\documentclass[fleqn,usenatbib]{mnras}

\pdfoutput=1 
\usepackage{newtxtext,newtxmath}
\usepackage{graphicx}	
\usepackage{nicefrac}
\usepackage{amsmath,amstext}
\usepackage[T1]{fontenc}
\DeclareRobustCommand{\VAN}[3]{#2}
\let\VANthebibliography\thebibliography
\def\thebibliography{\DeclareRobustCommand{\VAN}[3]{##3}\VANthebibliography}

\input{hyperlink-year-only-natbib-patch}

\newcommand*{\thethreehundred}{{\sc The Three Hundred}}
\newcommand*{\ltsim}{\ {\raise-.75ex\hbox{$\buildrel<\over\sim$}}\ }
\newcommand*{\gtsim}{\ {\raise-.75ex\hbox{$\buildrel>\over\sim$}}\ }

\newcommand*{\mysub}[2]{\ensuremath{#1_{\mathrm{#2}}}}
\newcommand*{\rs}{\mysub{r}{s}}

\graphicspath{{./}{figures/}}

\title[C--M Relation: Simulations vs Observations]{The Concentration--Mass Relation of Massive, Dynamically Relaxed Galaxy Clusters: Agreement Between Observations and $\Lambda$CDM Simulations}

\author[E. Darragh-Ford et al.]{%
Elise Darragh-Ford,$^{1,2,3}$\thanks{E-mail: \href{mailto:edarragh@stanford.edu}{edarragh@stanford.edu} (EDF)}
Adam B. Mantz,$^{1}$
Elena Rasia,$^{4, 5}$
Steven W. Allen,$^{1,2,3}$
R. Glenn Morris,$^{1,2}$\newauthor
Jack Foster,$^{3}$
Robert W. Schmidt,$^{6}$
and
Guillermo Wenrich$^{3}$
\smallskip
\\
$^{1}$Kavli Institute for Particle Astrophysics and Cosmology, Stanford University, 452 Lomita Mall, Stanford, CA 94305, USA\\
$^{2}$SLAC National Accelerator Laboratory, 2575 Sand Hill Road, Menlo Park, CA 94025, USA\\
$^{3}$Department of Physics, Stanford University, 382 Via Pueblo Mall, Stanford, CA 94305, USA\\
$^{4}$INAF - Osservatorio Astronomico di Trieste, via Tiepolo 11, I-34143 Trieste, Italy\\
$^{5}$ Institute of Fundamental Physics of the Universe, via Beirut 2, 34151 Grignano, Trieste, Italy \\
$^{6}$Astronomisches Rechen-Institut, Zentrum f\"ur Astronomie der Universit\"at Heidelberg, M\"onchhofstrasse 12-14, D-69120 Heidelberg, Germany
}

\date{Accepted XXX. Received YYY; in original form ZZZ}

\pubyear{2022}

\begin{document}
\label{firstpage}
\pagerange{\pageref{firstpage}--\pageref{lastpage}}
\maketitle

\begin{abstract}

The relationship linking a galaxy cluster's total mass with the concentration of its mass profile and its redshift is a fundamental prediction of the Cold Dark Matter (CDM) paradigm of cosmic structure formation.
However, confronting those predictions with observations is complicated by the fact that simulated clusters are not representative of observed samples where detailed mass profile constraints are possible.
In this work, we calculate the Symmetry-Peakiness-Alignment (SPA) morphology metrics  for maps of X-ray emissivity from \thethreehundred{} project hydrodynamical simulations of galaxy clusters at four redshifts, and thereby select a sample of morphologically relaxed, simulated clusters, using observational criteria. These clusters have on average earlier formation times than the full sample, confirming that they are both morphologically and dynamically more relaxed than typical. We constrain the concentration--mass--redshift relation of both the relaxed and complete sample of simulated clusters, assuming power-law dependences on mass ($\kappa_m$) and $1+z$ ($\kappa_\zeta$), finding $\kappa_m = -0.12 \pm 0.07$ and $\kappa_\zeta = -0.27 \pm 0.19$ for the relaxed subsample.
From an equivalently selected sample of massive, relaxed clusters observed with {\it Chandra}, we find $\kappa_m = -0.12 \pm 0.08$ and $\kappa_\zeta = -0.48 \pm 0.19$, in good agreement with the simulation predictions.
The simulated and observed samples also agree well on the average concentration at a pivot mass and redshift providing further validation of the $\Lambda$CDM paradigm in the properties of the largest gravitationally collapsed structures observed.
This also represents the first clear detection of decreasing concentration with redshift, a longstanding prediction of simulations, in data.
\end{abstract}

\begin{keywords}
galaxies: clusters: general -- X-rays: galaxies: clusters -- cosmology: observations
\end{keywords}

\section{Introduction} \label{sec:intro}

The $\Lambda$CDM paradigm has as one of its core pillars the assumption that the matter budget of the universe is dominated by dark matter in the form of cold, weakly interacting, massive particles. Under this paradigm we can make detailed predictions for the specific distribution of mass we expect to find within structures that form hierarchically, most immediately the relationship between total mass and the ``concentration'' parameter describing the distribution of mass within a halo \citep{Navarro9611107, Navarro2004, Gao2008}. Verifying these predictions is an important component of testing $\Lambda$CDM. However, observational tests require resolved measurements of the total density within dark-matter-dominated gravitational potentials, which can be difficult to obtain. The most common methods for obtaining such measurements are through gravitational lensing of background galaxies (e.g.\ \citealt{Okabe2013, Merten_2015,Du2015, Shan2017}) or X-ray observations of the intracluster medium, under the assumption of hydrostatic equilibrium (e.g.\ \citealt{Vikhlinin2006, Schmidt2007, Mantz_2016, Amodeo_2016}).
The approaches have complementary strengths and limitations: lensing is a more direct probe of gravitating mass, but is sensitive to all the mass projected along a line of sight, while X-ray measurements can more easily constrain the 3-dimensional mass profile but are reliable only for the most dynamically relaxed clusters due to the assumption of equilibrium.
In either case, care is needed to ensure that properties of measured cluster samples are compared with predictions for similarly selected samples from simulations as has been discussed previously in works such as \cite{Meneghetti_2014}.

In the case of X-ray measurements of dynamically relaxed clusters, the ideal case would be to compare with simulated clusters that satisfy an equivalent observational selection for dynamical relaxation based on mock-imaging information, particularly given that mock-observable proxies for relaxation have been found to correlate only weakly with typical dynamical proxies computed directly from simulated particles \citep{Cao_2021, De_Luca_2021}.
Typically, measurements of the morphology of the X-ray emitting gas are used as a proxy for dynamical relaxation in such a selection; broadly speaking, such measurements aim to distinguish relaxed clusters based on the sharpness of the surface brightness peak at the cluster center \citep{Vikhlinin2007, Santos2008} and/or the overall appearance of symmetry on larger scales \citep{Mohr1993, Buote1995, Jeltema2005, Nurgaliev2013, Rasia2013}.

In this paper, we aim to replicate a selection based on the SPA algorithm of \citet{Mantz15} on mock images of simulated clusters.
This algorithm in particular was developed to select relaxed clusters for cosmological tests based on the cluster gas-mass fraction ($f_\mathrm{gas}$), and was subsequently shown to produce a sample with smaller intrinsic scatter in $f_\mathrm{gas}$ than the previous ``by-eye'' selection \citep{Mantz14, Mantz22}, consistent with identifying more relaxed systems.

In this paper, we apply the SPA methodology to identify observationally relaxed cluster analogs over a range of redshifts from \thethreehundred{} project simulations \citep{Cui_2018}, a suite of 324 high-resolution hydrodynamical re-simulations of halos from the dark-matter-only MultiDark simulations.
We then determine the concentration--mass--redshift relation of this relaxed subset, and compare it with constraints from 44 hot, dynamically relaxed clusters observed by the {\it Chandra} X-ray telescope satisfying the same selection.
Section~2 provides a description of both the simulated and observed data sets. Section~3 describes the fitting procedure and gives results for the concentration--mass--redshift relation for both the simulated and observed samples with a brief comparison with literature results. 
We summarize our results in Section~\ref{sec:summary}.

\section{Data}

\subsection{Observed Cluster Sample}

\begin{figure}
  \centering
  \includegraphics[scale=0.5]{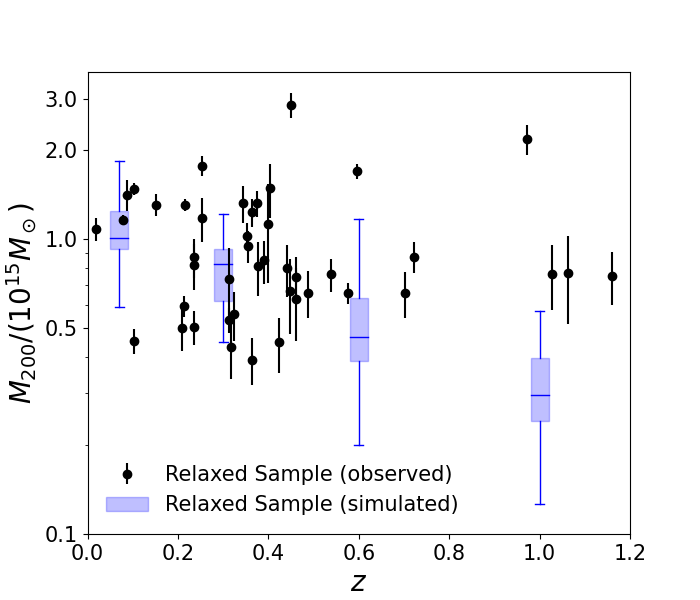}
  \caption{
    Mass--redshift for observed sample of 44 relaxed clusters (\emph{black points}) and relaxed simulated sample at each redshift (\emph{blue boxes}). The boxplots represent the [0, 0.25, 0.5, 0.75, 1.0] percentiles of relaxed clusters at each redshift slice. 
  }
\label{fig:m-Z}
\end{figure}

This work employs the sample of 44 hot, dynamically relaxed clusters observed by {\it Chandra} and analyzed by \cite{Mantz22}.
Each cluster is selected to have an X-ray morphology satisfying the SPA criteria for relaxation, and an ICM temperature $\geq 5$\,keV in the isothermal part of the profile ($\sim 0.5-1\,r_{2500}$).%
\footnote{We jointly define a family of characteristic masses and radii of clusters corresponding to ``overdensities'' $\Delta$ in the conventional way, with $M_\Delta = (4/3) \pi \Delta \rho_\mathrm{c}(z) r_\Delta^3$, where $\rho_\mathrm{c}(z)$ is the critical density at a cluster's redshift.}
The sample spans redshifts $0.018 < z < 1.16$. The mass--redshift distribution of the sample can be seen in Figure~\ref{fig:m-Z} (\emph{black points}). 
The analysis of \citet[][see also \citealt{Mantz14, Mantz16}]{Mantz22} provides constraints on a parametrized NFW (\citealt*{Navarro9611107}) density profile for each cluster,
\begin{equation}
  \rho(r) \propto \left(\frac{r}{\rs}\right)^{-1} \left(1 + \frac{r}{\rs}\right)^{-2},
\end{equation}
under the assumptions of spherical symmetry and hydrostatic equilibrium.
In brief, X-ray spectra extracted from concentric annuli are used to simultaneously fit the NFW mass profile and a non-parametric temperature profile that captures the temperature decrease towards the center of a cool-core cluster.
With small variations due to varying cluster morphologies and data quality, spectral information sufficient to constrain temperatures -- and thus the mass profile -- is used at radii from $\sim50$\,kpc to $>r_{2500}$, which for typical concentrations spans the NFW scale radius, $\rs$.
The data thus provide information about the normalization of the mass profile and its shape, which can be expressed as joint constraints (including correlation) on $M_{200}$ and the associated concentration, $c = r_{200}/\rs$, shown in Table~\ref{tab:cM}.
The adequacy of the NFW model is verified by comparing the density and temperature profiles with those obtained from non-parametric fits (i.e., not assuming any particular mass model; see e.g.\ the appendix of \citealt{Mantz16}).
For further details, we refer the reader to \citet{Mantz14, Mantz16, Mantz22}.

\begin{table*}
  \centering
  \caption{
    Redshifts, masses and concentrations of clusters in our observed sample from X-ray data.
    The final column provides the correlation coefficient between measurement uncertainties in $\ln M_{200}$ and $\ln c$, which is accounted for in further analysis.
  }
  \vspace{1ex}
  \begin{tabular}{lcr@{ $\pm$ }lr@{ $\pm$ }lc}
    \hline
    Cluster & $z$ & \multicolumn{2}{c}{$M_{200}$ $(10^{15}M_\odot)$} & \multicolumn{2}{c}{$c$} & $\rho_{\ln}$\\
        \hline\vspace{-3ex}\\
    Perseus~Cluster      &  0.018  &  1.08  &  0.10  &  4.61  &  0.33  &  -0.76  \\
Abell~2029           &  0.078  &  1.16  &  0.04  &  6.36  &  0.21  &  -0.95  \\
Abell~478            &  0.088  &  1.41  &  0.17  &  4.91  &  0.48  &  -0.94  \\
PKS~0745$-$191       &  0.103  &  1.48  &  0.07  &  5.20  &  0.18  &  -0.93  \\
RX~J1524.2$-$3154    &  0.103  &  0.45  &  0.04  &  7.81  &  0.81  &  -0.91  \\
Abell~2204           &  0.152  &  1.31  &  0.11  &  7.68  &  0.60  &  -0.92  \\
RX~J0439.0+0520      &  0.208  &  0.50  &  0.08  &  8.07  &  1.50  &  -0.92  \\
Zwicky~2701          &  0.214  &  0.59  &  0.05  &  5.41  &  0.42  &  -0.92  \\
RX~J1504.1$-$0248    &  0.215  &  1.31  &  0.06  &  7.28  &  0.25  &  -0.86  \\
Zwicky~2089          &  0.235  &  0.50  &  0.07  &  5.60  &  0.65  &  -0.93  \\
RX~J2129.6+0005      &  0.235  &  0.82  &  0.14  &  6.29  &  1.52  &  -0.92  \\
RX~J1459.4$-$1811    &  0.236  &  0.87  &  0.13  &  5.38  &  1.14  &  -0.82  \\
Abell~1835           &  0.252  &  1.78  &  0.14  &  4.55  &  0.37  &  -0.94  \\
Abell~3444           &  0.253  &  1.18  &  0.20  &  4.17  &  0.73  &  -0.94  \\
MS~2137.3$-$2353     &  0.313  &  0.53  &  0.05  &  7.91  &  0.82  &  -0.90  \\
MACS~J0242.5$-$2132  &  0.314  &  0.73  &  0.21  &  7.82  &  2.32  &  -0.83  \\
MACS~J1427.6$-$2521  &  0.318  &  0.43  &  0.09  &  8.94  &  2.46  &  -0.87  \\
MACS~J2229.7$-$2755  &  0.324  &  0.56  &  0.11  &  6.50  &  1.28  &  -0.90  \\
MACS~J0947.2+7623    &  0.345  &  1.32  &  0.19  &  5.88  &  0.67  &  -0.93  \\
MACS~J1931.8$-$2634  &  0.352  &  1.03  &  0.10  &  5.87  &  0.59  &  -0.92  \\
MACS~J1115.8+0129    &  0.355  &  0.95  &  0.12  &  6.50  &  0.97  &  -0.84  \\
MACS~J1532.8+3021    &  0.363  &  1.24  &  0.13  &  4.79  &  0.34  &  -0.95  \\
MACS~J0150.3$-$1005  &  0.363  &  0.39  &  0.07  &  7.66  &  1.73  &  -0.78  \\
RCS~J1447+0828       &  0.376  &  1.32  &  0.14  &  5.00  &  0.34  &  -0.85  \\
MACS~J0011.7$-$1523  &  0.378  &  0.81  &  0.17  &  7.02  &  1.87  &  -0.90  \\
MACS~J1720.2+3536    &  0.391  &  0.85  &  0.14  &  6.77  &  1.51  &  -0.73  \\
MACS~J0429.6$-$0253  &  0.399  &  1.12  &  0.41  &  4.58  &  1.41  &  -0.91  \\
MACS~J0159.8$-$0849  &  0.404  &  1.49  &  0.31  &  5.75  &  1.54  &  -0.89  \\
MACS~J2046.0$-$3430  &  0.423  &  0.45  &  0.09  &  7.31  &  1.60  &  -0.92  \\
IRAS~09104+4109      &  0.442  &  0.80  &  0.16  &  6.38  &  1.11  &  -0.92  \\
MACS~J1359.1$-$1929  &  0.447  &  0.67  &  0.19  &  5.73  &  1.50  &  -0.91  \\
RX~J1347.5$-$1145    &  0.451  &  2.86  &  0.29  &  7.49  &  0.52  &  -0.87  \\
3C~295               &  0.460  &  0.62  &  0.17  &  5.85  &  1.37  &  -0.94  \\
MACS~J1621.3+3810    &  0.461  &  0.74  &  0.13  &  6.98  &  1.52  &  -0.89  \\
MACS~J1427.2+4407    &  0.487  &  0.66  &  0.12  &  6.74  &  1.37  &  -0.77  \\
MACS~J1423.8+2404    &  0.539  &  0.76  &  0.10  &  7.05  &  0.90  &  -0.89  \\
SPT~J2331$-$5051     &  0.576  &  0.66  &  0.05  &  5.34  &  0.29  &  -0.32  \\
SPT~J2344$-$4242     &  0.596  &  1.70  &  0.10  &  6.57  &  0.27  &  -0.51  \\
SPT~J0000$-$5748     &  0.702  &  0.66  &  0.12  &  5.16  &  0.75  &  -0.86  \\
SPT~J2043$-$5035     &  0.723  &  0.87  &  0.10  &  3.90  &  0.25  &  -0.81  \\
SPT~J0615$-$5746     &  0.972  &  2.19  &  0.26  &  3.26  &  0.33  &  -0.77  \\
CL~J1415.2+3612      &  1.028  &  0.76  &  0.19  &  3.86  &  0.71  &  -0.90  \\
3C~186               &  1.063  &  0.77  &  0.25  &  3.80  &  1.00  &  -0.90  \\
SPT~J2215$-$3537     &  1.160  &  0.75  &  0.15  &  5.40  &  0.70  &  -0.83  \\

    \hline
    \end{tabular}
  \label{tab:cM}
\end{table*}

All cosmology-dependent calculations assume a flat $\Lambda$CDM model with Hubble constant $H_0=67.8$ km/s/Mpc and mean matter density $\Omega_\mathrm{m}=0.307$, for consistency with the simulations (see below).

\begin{figure*}
\centering
\includegraphics[width=0.3\textwidth]{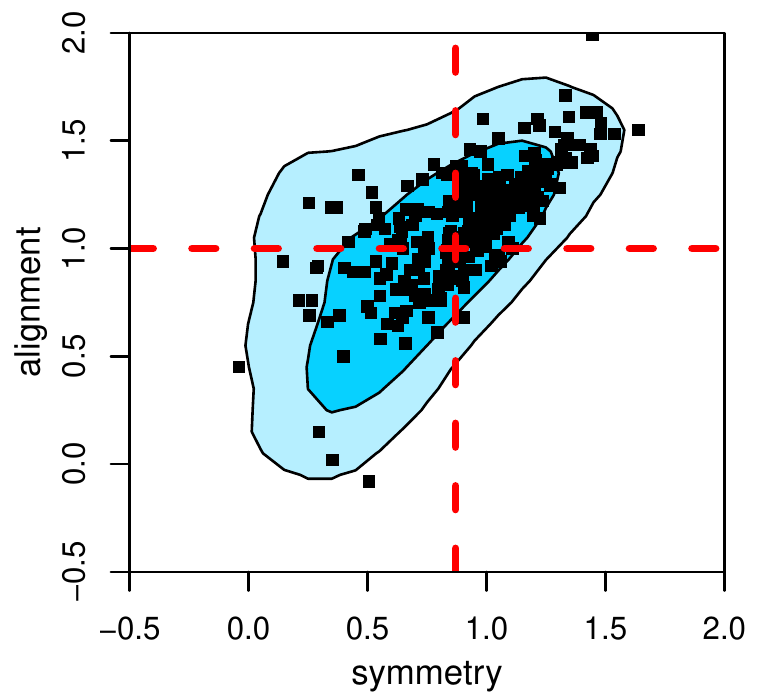}
\includegraphics[width=0.3\textwidth]{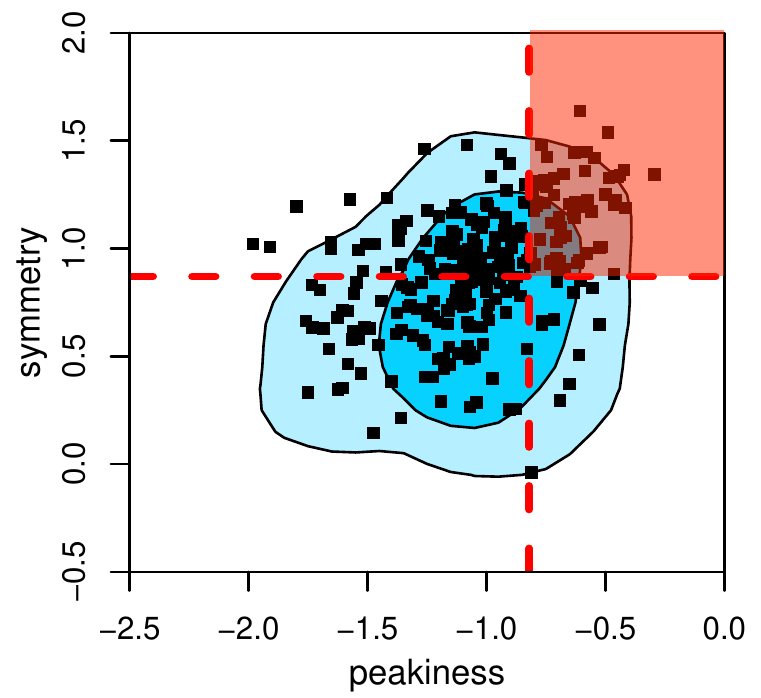}
\includegraphics[width=0.3\textwidth]{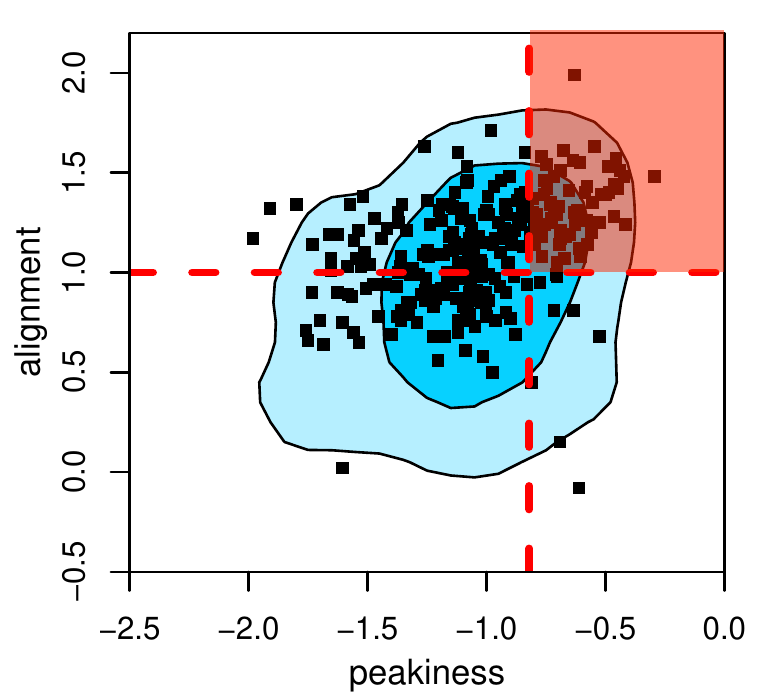}
\caption{
    Pairwise comparison of the distributions of the 3 SPA metrics found from simulated X-ray maps (blue; contours containing 68.3\% and 95.4\% of projections) with observed data from \citet[black points]{Mantz15}.
    Red lines mark the threshold values that must all be exceeded for a cluster to be considered relaxed.}

\label{fig:s-p-a}
\end{figure*}

\subsection{Simulated Cluster Sample} \label{sec:sims}

\begin{figure*}
\centering
\includegraphics[width=0.85\textwidth]{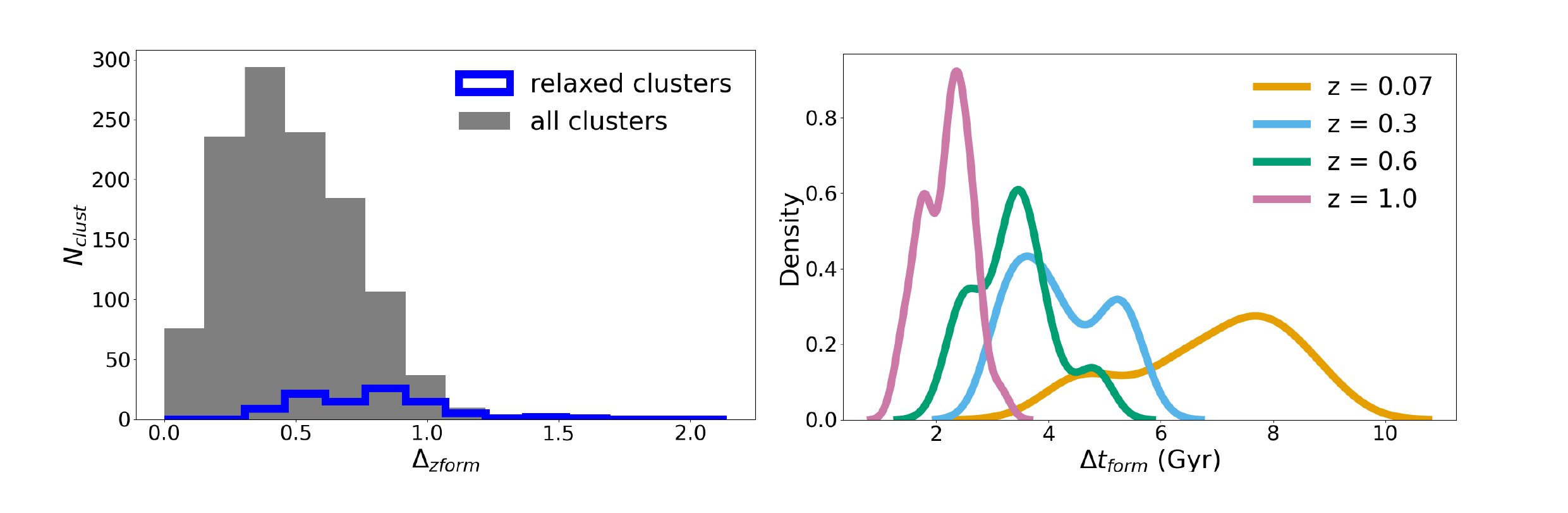}
\caption{\emph{left:} Histograms showing  distribution of $\Delta z_\mathrm{form}$ for the complete simulated sample (\emph{black}) and for the relaxed simulated sample only (\emph{blue}). \emph{right:} Kernel density estimate (KDE) plot of the distribution of time since formation for the samples of relaxed simulated clusters at each redshift. }
\label{fig:delta_z}
\end{figure*}

The simulated cluster sample is selected from \thethreehundred{} project, which is a set of 324 hydrodynamical re-simulations of the most massive clusters (at $z=0$) from the (1 Gpc/h)$^{3}$ dark-matter-only MDPL2 MultiDark simulation \citep{Klypin2016}.

The SPA selection is based on X-ray emissivity maps generated at four redshifts ($z = [0.067, 0.333, 0.592, 0.986]$), approximately spanning the redshift range of the observed sample. 

For each re-simulation box at each redshift, the map-making procedure identifies the most massive halo in the box (which is usually but not always the same halo at each redshift; \citealt{Ansarifard1911.07878}).
Maps of X-ray emissivity in the rest frame energy band 0.5--2.0 keV, integrated along the line of sight, in three orthogonal projections, are produced with the code \emph{Smac} \citep{Doulag2005}. The maps include particles with temperatures above $10^6$\,K and densities below the density for star formation. Each particle is weighted by a spline kernel of width equal to the gas particle smoothing length. All maps are $4\times4$ Mpc and $1024\times1024$ pixels, leading to a physical resolution of 3.9 kpc per pixel. The integration length along the line of sight is 10\,Mpc. Each of the maps are analyzed separately for a total of $\sim 900$ projections at each redshift. 
Note that these maps contain neither mock-observational artifacts nor emissive sources other than the ICM, though they do contain emission from multiple halos in each simulation box, if present.
Since our intention is to quantify the intrinsic morphology of the target cluster, emissive structures that are unambiguously projections (those appearing in only one projection, since this requires them to be $>4$\,Mpc distant from the cluster center) were masked in the analysis below.

We computed the SPA metrics independently for each projection of each simulated cluster, as detailed in Appendix~\ref{sec:spametrics}.
Figure~\ref{fig:s-p-a} shows that the distributions of the 3 metrics estimated by the SPA algorithm\footnote{For a detailed definition of the SPA metrics see Appendix~\ref{sec:spametrics}.} -- symmetry ($s$), peakiness ($p$) and alignment ($a$) -- from the simulations are broadly compatible with the sample of 361 observed clusters analyzed by \cite{Mantz15}, although in detail the distributions differ somewhat (Appendix~\ref{sec:appendix_results}). 
Approximately 8\% of projections at each redshift satisfy the SPA criteria for relaxation, with a weak trend in the selected fraction with redshift. This is similar to the relaxed fraction found by \cite{Mantz15} for clusters selected from Sunyaev-Zel'dovich effect surveys; this is a more fair comparison than with X-ray selected clusters, as peaky clusters (possessing cool, X-ray bright cores) are preferentially detected in X-ray surveys \citep{Rossetti_2017}.
We consider a simulated cluster to be relaxed if at least two of its projected maps satisfy the SPA criteria. Of all of the clusters, $87.7\%$ are considered relaxed in zero projections, 4.6\% are considered relaxed in one projection, 2.4\% are considered relaxed in two projections, and 5.3\% are considered relaxed in all three projections. Thus, similarly to the individual projections, approximately 8\% of the simulated clusters overall are identified as relaxed, with a weak trend in the selected fraction with redshift (Appendix~\ref{sec:appendix_results}). 

A comparison of the mass--redshift distribution for the simulated (\emph{blue boxes}) and observed (\emph{black points}) relaxed cluster samples can be seen in Figure~\ref{fig:m-Z}. The two samples cover a similar range in mass and redshift at most redshifts. However, the observed sample at high redshifts ($z > 0.8)$ are all at higher masses than the simulated sample. This is likely due to the large available observational volume relative to the simulated sample, combined with the difficulties of observing lower mass objects at the highest redshifts. 

A simple way to quantify the dynamical history of a simulated cluster is through the time since formation (difference in lookback time between formation and observation; $\Delta t_\mathrm{form}$), with ``formation'' conventionally defined as the point when half of the clusters mass at the time of observation had been accumulated (e.g.\ \citealt{Ludlow1206.1049}); clusters with longer times since formation should be more relaxed.
We find that the SPA algorithm does indeed preferentially select clusters with longer times since formation.
In practice, the distribution of $\Delta t_\mathrm{form}$ (regardless of selection) varies with the redshift of observation, whereas the distribution of differences in redshift between formation and observation, $\Delta z_\mathrm{form}$, is relatively static for both the SPA-selected and unselected clusters.
The latter thus provides a more straightforward way of quantifying the difference between the two samples, as shown in the left panel of Figure~\ref{fig:delta_z}. 
On average, the SPA-selected sample has a higher value of $\Delta z_\mathrm{form}=0.75 \pm 0.28$ compared with $0.46 \pm 0.25$ for the unrelaxed sample; a Kolmogorov-Smirnoff test rejects the hypothesis that the two samples are equivalent with a p-value of $<10^{-16}$.
The right panel of Figure~\ref{fig:delta_z} shows the distributions of $\Delta t_\mathrm{form}$ for the SPA-selected clusters in each redshift slice.

For all the simulated clusters, values of $r_{200}$ and $M_{200}$ were determined directly from 3D mass profiles generated about each cluster center.
Concentrations were estimated by fitting NFW models to the 3D profiles between radii of $0.05\,r_{200}$ and $r_{200}$ (following \citealt{Meneghetti_2014}), minimizing the difference in log-mass between the model and data, with equal weighting.
For the relaxed selection, we expect higher than average concentrations (corresponding to earlier formation times, i.e.\ longer times since the last major merger), as the next section verifies.

\begin{figure*}
  \centering
  \includegraphics[scale=0.9]{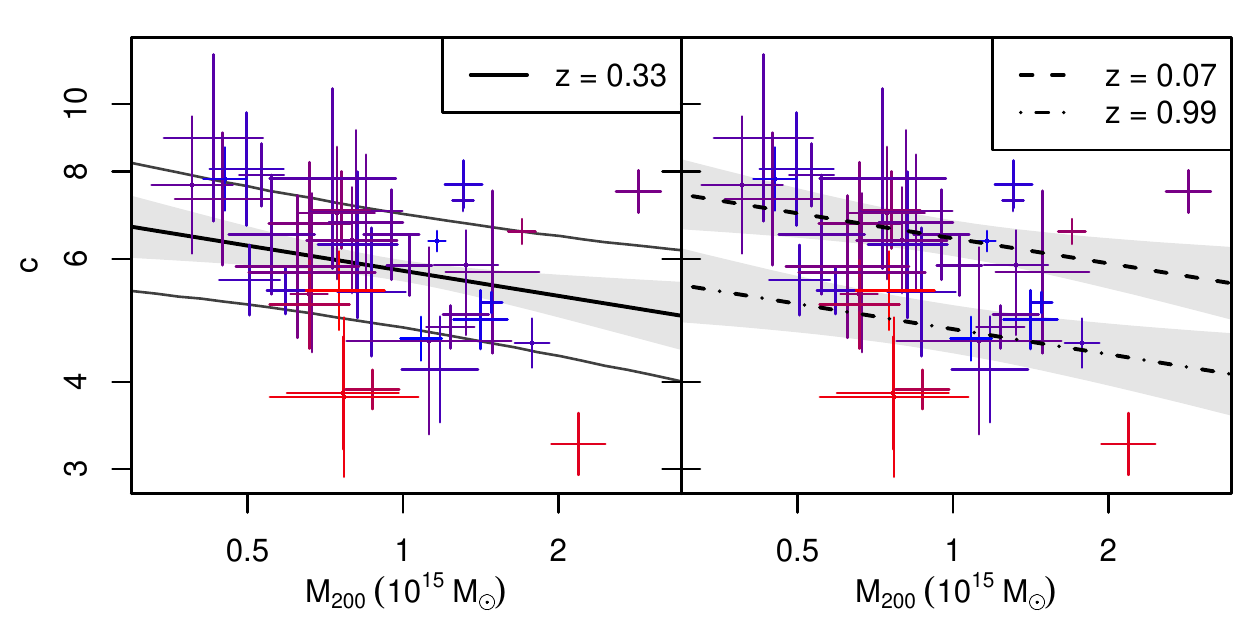}
  \includegraphics[scale=0.9]{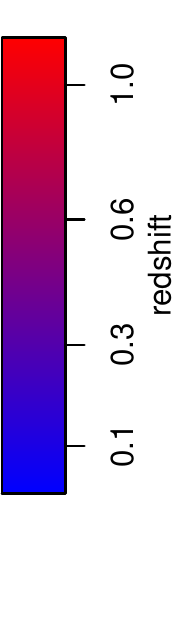}
  \caption{
    Left: Mass--concentration data for the observed sample of 44 relaxed clusters, color coded from blue to red by increasing redshift. Though not shown here, the correlation between mass and concentration measurements is accounted for in our analysis. The dark curve shows the best-fitting power-law with mass at approximately the median redshift of the data, with the shaded region indicating the (68.3\%) statistical uncertainty in the fit. The lighter curves delineate the 68.3\% probability predictive interval, which includes the impact of the intrinsic scatter.
    Right: the same data are compared with the model predictions and their statistical uncertainty at redshifts of 0.067 and 0.986. 
  }
\label{fig:chandra}
\end{figure*}

\section{Concentration--Mass--Redshift Relation}
\label{c-m}

Following \citep{Mantz_2016} (hereafter M16), we consider a power-law model for concentration as a function of both mass and redshift:
\begin{equation}
\label{eq:c-m}
  \ln\left(\frac{c}{c_0}\right) = \kappa_\zeta \ln{\left(\frac{1+z}{1.35} \right)} + \kappa_m \ln{\left (\frac{M_{200}}{10^{15} M_\odot} \right)}.
\end{equation}

We use the {\sc lrgs} linear regression code\footnote{\url{https://github.com/abmantz/lrgs}} to fit this model to both the observed and simulated cluster data, approximating the uncertainty in masses and concentrations determined from the X-ray data as jointly log-normal and accounting for its covariance.
We do not include any uncertainty on masses and concentrations from the simulations. 
The model includes a log-normal intrinsic scatter, $\sigma$, which is assumed to be constant with mass and redshift.

\begin{figure}
  \centering

    \includegraphics[width=0.35\textwidth]{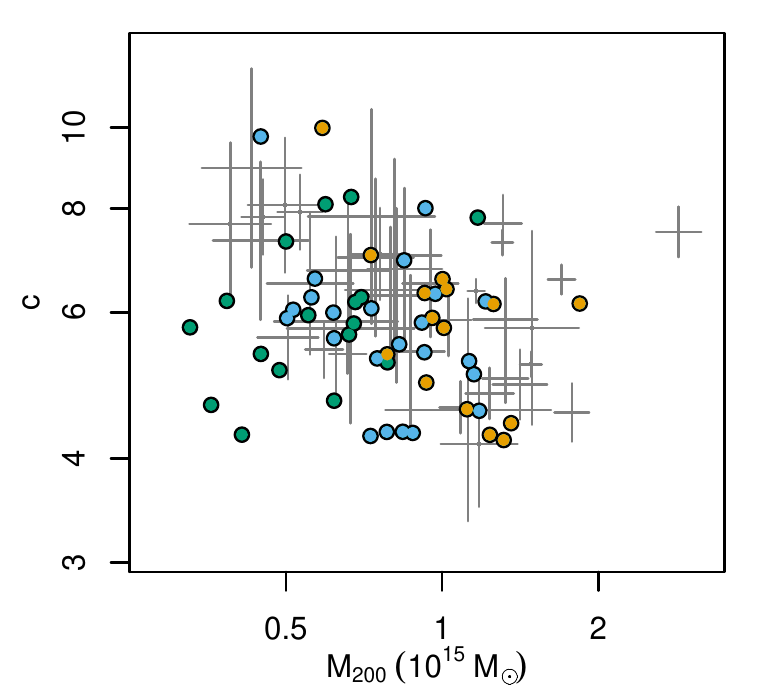}

  \caption{
    The observed data (\emph{grey}) restricted to $z < 0.6$ compared with the relaxed clusters from the simulated data at z = 0.07 (orange), 0.3 (blue), and 0.6 (green). 
  }
\label{fig:chandra_sim}
\end{figure}

\begin{figure*}
\centering
\includegraphics[width=0.45\textwidth]{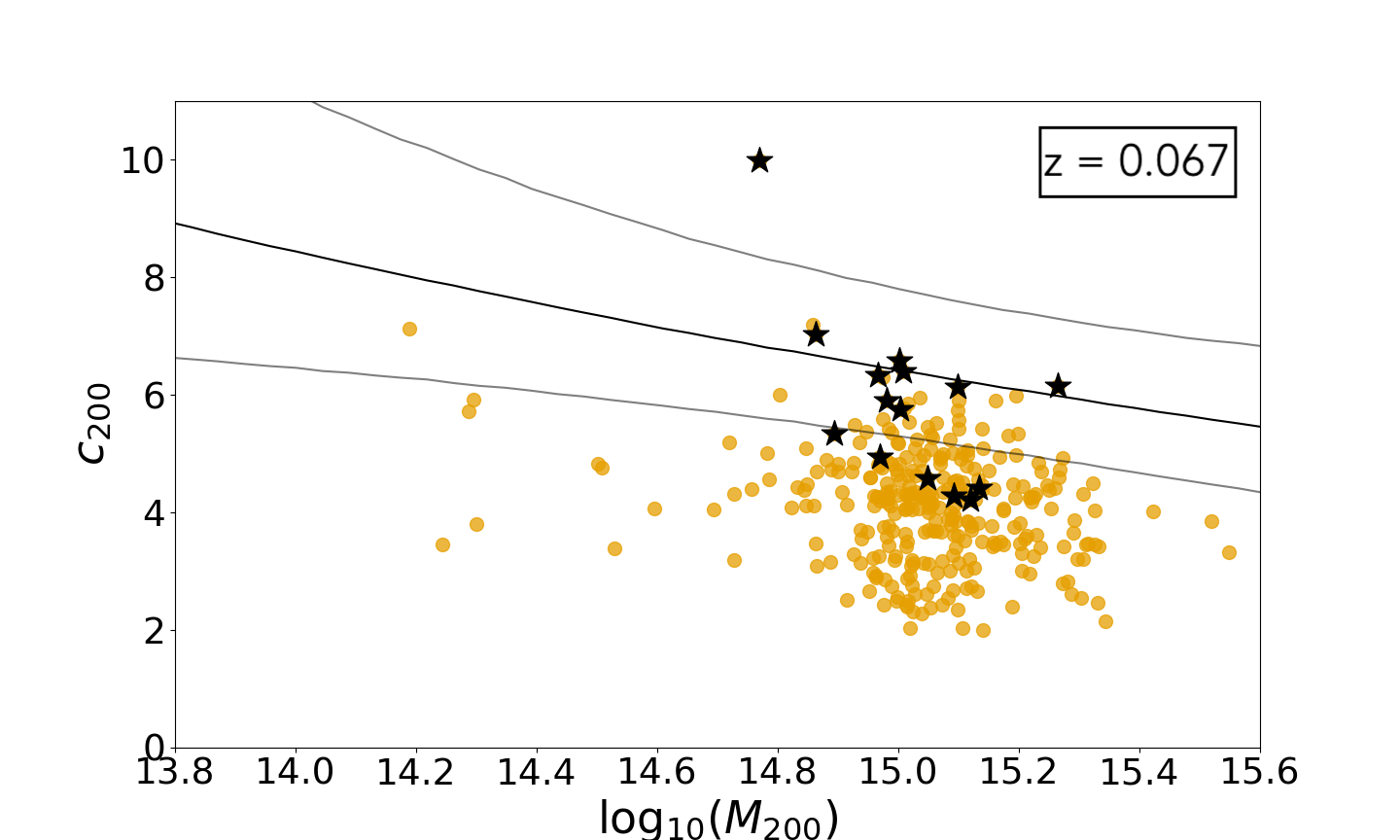}
\includegraphics[width=0.45\textwidth]{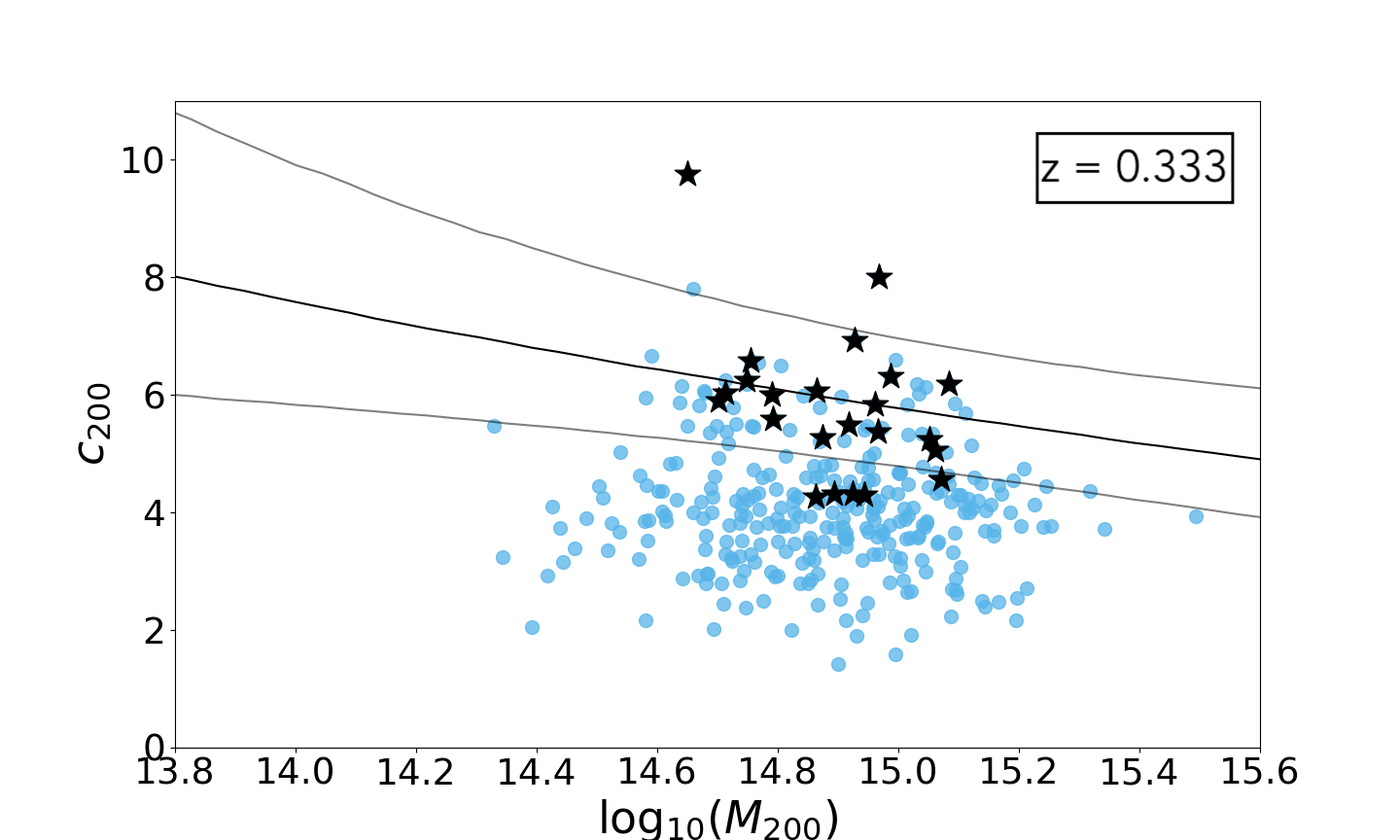}
\includegraphics[width=0.45\textwidth]{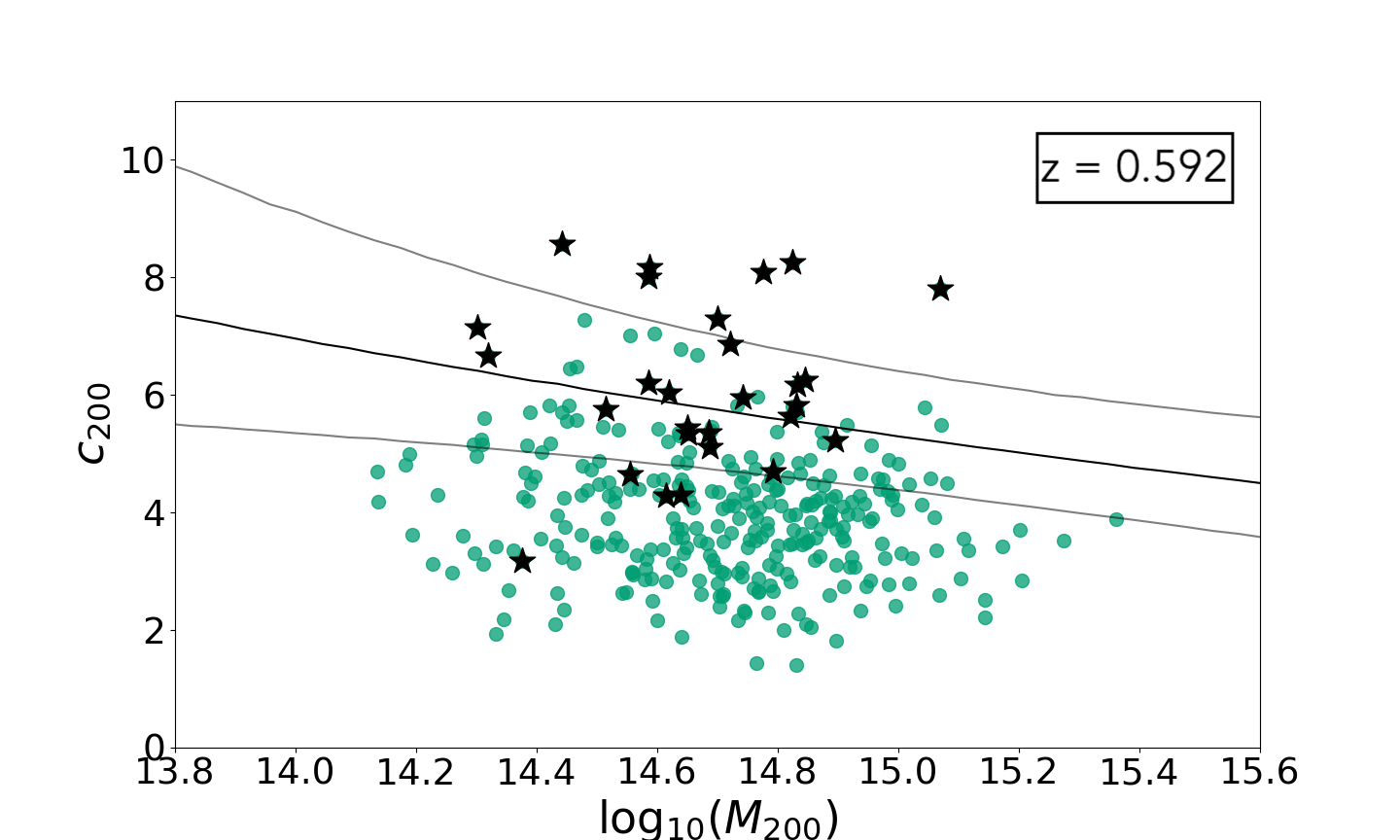}
\includegraphics[width=0.45\textwidth]{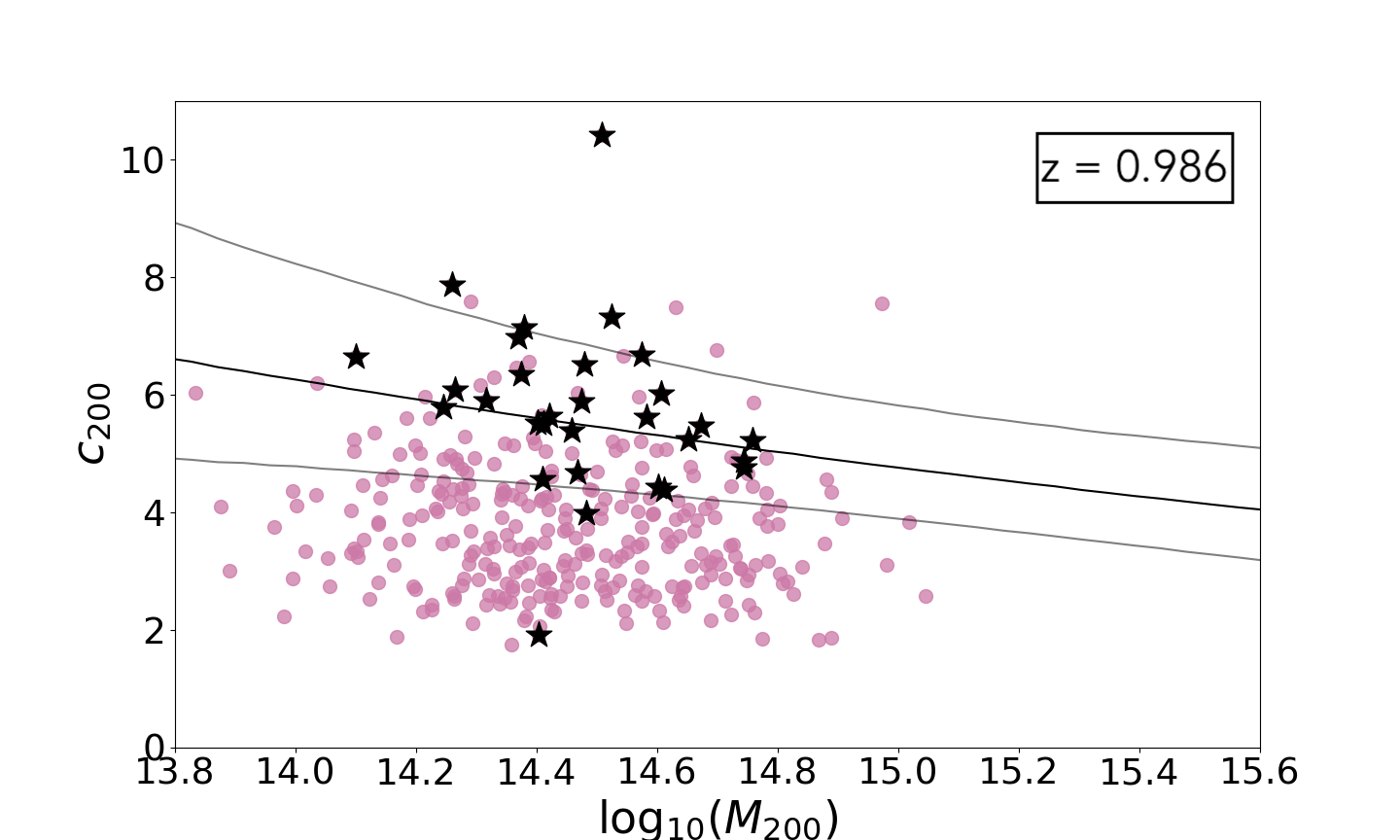}
\caption{Mass--concentration relation of simulated clusters as a function of redshift (\emph{top left}: z = 0.067,  \emph{top right}: z = 0.333, \emph{bottom left}: z = 0.592, and \emph{bottom right}: z = 0.986). SPA-selected (i.e. relaxed) clusters from the simulations are shown as filled black stars. The black line is the best-fitting model from the analysis of the observed data at each  redshift, with the lighter curves showing the predictive interval (accounting for intrinsic scatter). There is good agreement between the relaxed simulated data and the fits to the relaxed observed clusters.} 
\label{fig:mass_conc}
\end{figure*}

\begin{table*}
\centering
\caption{Mass--concentration relation fit to different samples of clusters. \emph{Rows 1--4} correspond simulations and observations analyzed in this work, as well as the previous generation of observed SPA-selected clusters. \emph{Rows 5--7} show fits to other simulated cluster samples. \emph{Rows 8--9} show independent measurements in the recent literature from X-ray and gravitational lensing data. The text provides more details about the different sample selections.
  Uncertainties on the $c_0$ values derived from the literature should be considered approximate, since we have no way of accounting for the correlations between parameter uncertainties from those fits.
  For the Dutton \& Maccio work, we have estimated $\kappa_\zeta$ by fitting a power law to the normalization parameters of their fits at $z=0$, 0.5 and 1.0.
}
\label{tab:c-m_fit} 
    \begin{tabular}{lccccr}
    \hline

    Cluster sample & Selection & $\kappa_m$ & $\kappa_\zeta$ & $c_0$ & $\sigma$ \\

    \hline
    \thethreehundred{} Simulations & all clusters &  $-0.06 \pm 0.02$ & $-0.31 \pm 0.07$ & $4.31 \pm 0.13$ & $0.30 \pm 0.01$ \\
    \thethreehundred{} Simulations & SPA &  $-0.12 \pm 0.07$ & $-0.27 \pm 0.19$ & $5.99 \pm 0.35
$ & $0.24 \pm 0.02$ \\
    {\it Chandra} Data (this work) & SPA &  $-0.12 \pm 0.08$ & $-0.48 \pm 0.19$ & $5.74 \pm 0.20$ & $0.18 \pm 0.03$ \\
    {\it Chandra} Data \citep{Mantz_2016} & SPA &  $-0.16 \pm 0.07$ & $-0.17 \pm 0.26$ & $6.01 \pm 0.23$ & $0.16 \pm 0.03$ \\
    \hline
    \texttt{MUSIC-2} Simulations \citep{Meneghetti_2014} & CLASH  &  $-0.08 \pm 0.02$ & $-0.20 \pm 0.09$ & $4.10 \pm 0.07$ & 0.16 \\
    \emph{Planck} Cosmology Simulations \citep{Dutton_2014} &$\Delta r$ and $\rho_{\rm rms}$ & $-0.09 \pm 0.001$ & $-0.58 \pm 0.05$ & $3.72 \pm 0.009$ & 0.11 \\
    ``Q-Continuum/Outer Rim'' Simulations \citep{Child_2018}  & $\Delta r$ &  $-0.08$ & $-0.35$ & $3.51$ & -- \\
    Strong+Weak Gravitational Lensing Data \citep{Merten_2015}  &CLASH  & $-0.32 \pm 0.18$ &  $0.14 \pm 0.52$ & $3.85 \pm 0.17$ & 0.07 \\
    {\it Chandra} Data \citep{Amodeo_2016}  & by-eye &  $-0.50 \pm 0.20$ & $0.12 \pm 0.61$ & $4.63 \pm 3.09$ & $0.14 \pm 0.09$ \\
    \hline
\end{tabular}
     
\end{table*}

Figure~\ref{fig:chandra} (left panel) shows the observed clusters, along with the best-fitting mass--concentration relation at $z=0.33$ (approximately the sample median).
Table~\ref{tab:c-m_fit} lists the individual parameter constraints along with the previous results from M16, based on an earlier, slightly smaller, version of the observed SPA cluster sample.
Compared with the M16 sample, the current data incorporate precise measurements of the Perseus Cluster at $z=0.018$, additional relaxed clusters at $z=0.376$, 0.972 and 1.160, and numerous re-observations of previously included clusters (see \citealt{Mantz22} for details).
Owing to this increase in precision at the lowest and highest redshifts, our updated analysis finds significant evidence of evolution ($\kappa_\zeta \neq 0$), unlike the M16 analysis which showed an evolution consistent with zero; in contrast the constraints on $\kappa_m$ and $\sigma$ are similar to M16.
The right panel of Figure~\ref{fig:chandra} compares the data with the model predictions at low and high redshifts, providing a visual indication of the strength of the observed evolution. 

Figure~\ref{fig:chandra_sim} shows a comparison between the observed data and the relaxed clusters in the simulations.  The average concentration and evolution with mass show good visual agreement. This agreement is also visible in Figure~\ref{fig:mass_conc}, which shows the mass--concentration data from the simulations, highlighting the relaxed simulated clusters, along with the model fits to the observed data. It is clear again that the fit to the observed data shows qualitative agreement with the simulated sample of relaxed clusters. As expected, the SPA-selected, relaxed clusters (both simulated and observed) have on average higher concentrations at a given mass than the full simulated sample.

Table~\ref{tab:c-m_fit} shows the parameter constraints from fitting the functional form for Equation~\ref{eq:c-m} to the simulated data (both the relaxed subset and all clusters). As expected, the two fits differ significantly in the normalization parameter, with the typical relaxed cluster having a concentration of 5.8, compared with 4.0 for the average cluster. Interestingly, the constraints on the power-law indices with mass and redshift, and the scatter, are consistent between the relaxed subsample and all clusters in the simulated data set.  

\begin{figure}
\centering
\includegraphics[width=\columnwidth]{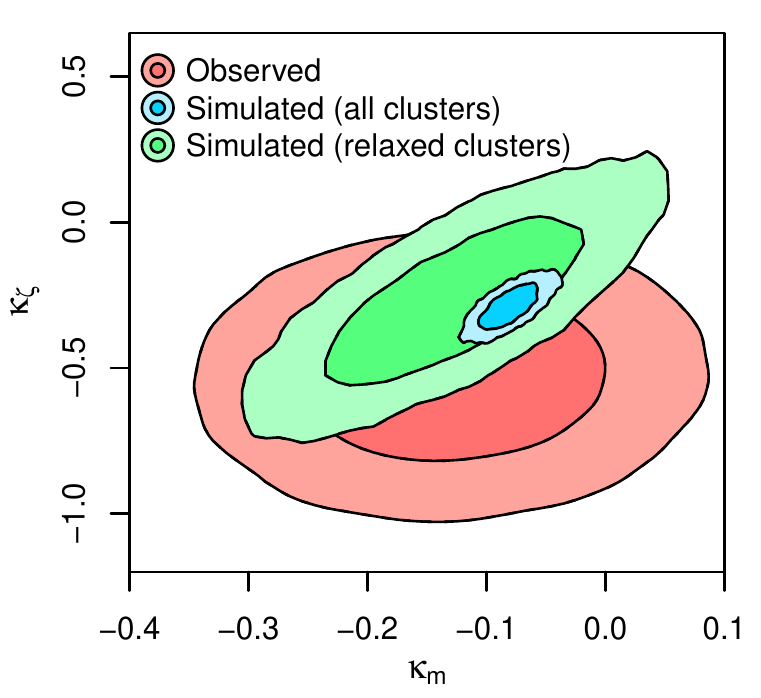}
\caption{Credible regions showing the degeneracy between power-law indices for mass ($\kappa_m$) and redshift ($\kappa_\zeta$) dependence of the concentration. \emph{blue:} shows the fit including all clusters in the simulation, \emph{green:} shows the fit only including relaxed simulated clusters, and \emph{red:} shows the fit to the observational data for relaxed clusters.}
\label{fig:fit_contours}
\end{figure}

Figure~\ref{fig:fit_contours} compares the joint constraints on the mass and redshift dependence from the fits to the relaxed observed clusters, relaxed simulated clusters, and all simulated clusters. The degeneracy between these parameters in the case of the simulated samples is a natural consequence of the mass distribution as a function of redshift of the simulated halos, visible in Figures~\ref{fig:m-Z} and \ref{fig:mass_conc}.
The agreement between the three results is excellent, with each be consistent with the others within the 68.3 per cent credible region. The normalization parameter from the observed clusters is similarly in good agreement with that from the relaxed simulations, and both are higher than the normalization fitted to all simulated clusters, as expected (Table~\ref{tab:c-m_fit}). We note that intrinsic scatter of both the relaxed subset and all simulated clusters is in excess of that inferred from the observed data.
The reason for this is not immediately clear, though we have verified that is not solely due to the few clear outliers in Figure~\ref{fig:mass_conc}. We check whether the radius used to measure the concentration in the simulations is significant by repeating the fit with concentrations measured out to $r_{500}$ rather than $r_{200}$. While we see a slight increase in the median concentrations at each redshift, we find that the slopes as a function of mass and redshift remain consistent within $1\sigma$. 

We see good agreement with previous estimates of the concentration--mass--redshift relation from both simulations \citep{Meneghetti_2014, Dutton_2014, Child_2018} and observational studies of X-ray selected clusters \citep{Amodeo_2016} and weak lensing analyses \citep{Merten_2015} (Table \ref{tab:c-m_fit}). Each of these works apply a different criteria for selecting relaxed clusters. \cite{Child_2018} selects relaxed clusters between $0 \leq z \leq 1$ based on a cut on distance between the halo center and center of mass ($\Delta r$). \cite{Dutton_2014} also selects clusters using a cut on $\Delta r$, but applies an additional cut based on the goodness of the fit of an NFW profile ($\rho_{\rm rms}$). \cite{Amodeo_2016} selected clusters from {\it Chandra}  between $0.4 < z < 1.2$ after removing clusters with clear evidence of dynamic activity by eye. Lastly \cite{Merten_2015} uses the 25 massive clusters between between $0.19 < z < 0.89$ in the CLASH dataset, 20 of which were chosen by eye for their relaxed appearance in X-ray data, while the sample in \cite{Meneghetti_2014} is explicitly chosen from the \texttt{MUSIC-2} Simulations to match the mass and other X-ray morphology metric distributions of the CLASH sample. However, despite different selection criteria, all of the simulated studies find a dependence on mass with a power-law index of $\sim -0.1$. The observational studies generally find a slightly stronger mass dependence, but agree with the simulation predictions within $\sim2\sigma$. The slope with redshift shows more variation, with  \cite{Amodeo_2016} and \cite{Merten_2015} finding a slope consistent with no or even positive redshift dependence, while \cite{Meneghetti_2014} and \cite{Child_2018} prefer a negative slope, in agreement with our findings. However, all of these results are consistent with ours within their relatively large uncertainties. We see more variation in the normalization of the various fits, likely being driven by differences in the selection criteria for each study. 
 
\section{Summary} \label{sec:summary}

We measured the galaxy cluster concentration--mass--redshift relation from {\it Chandra} data for a sample of 44 massive, relaxed clusters, as well as for a set of relaxed clusters from \thethreehundred{} project simulated data set at $z = [0.067, 0.333, 0.592, 0.996]$. We applied equivalent selection criteria in both cases, using the SPA X-ray morphology algorithm \citep{Mantz15}. We find excellent agreement between the observed data and the predictions for similarly relaxed clusters from $\Lambda$CDM simulations in the average concentration, as well as its dependence on mass and redshift. These results provide powerful further validation of the $\Lambda$CDM paradigm in the properties of the largest gravitationally collapsed structures observed. We find, for the first time, evidence for evolution in the mean concentration from the observed data, with concentrations nominally decreasing by 28 per cent between redshifts 0 and 1. Such evolution is also clearly present in the full simulated sample, although the simulated relaxed subsample is consistent with no evolution at the $2\sigma$ level due to its smaller size (and thus larger uncertainties). Constraints from the relaxed simulated clusters are broadly comparable in power to those from the observed data. Particularly as more relaxed clusters are found and observed, larger suites of hydrodynamical simulations (with mock images and appropriate selection) will be needed to provide precise predictions for comparison with observations.

\section*{Acknowledgements}
We thank Risa Wechsler for useful discussion, and Benjamin T. Floyd, Taweewat Somboonpanyakul, and Michael McDonald for helpful comments on the draft.

The scientific results reported in this article are based on observations made by the Chandra X-ray Observatory, and data obtained from the Chandra Data Archive.
Support for this work was provided by the National Aeronautics and Space Administration through Chandra Awards Number GO5-16122X, GO5-16148X, GO6-17112B, GO0-21124B and AR9-20012X issued by the Chandra X-ray Center, which is operated by the Smithsonian Astrophysical Observatory for and on behalf of the National Aeronautics Space Administration under contract NAS8-03060.
This research has made use of software provided by the Chandra X-ray Center (CXC) in the application packages CIAO.
We acknowledge support from the U.S. Department of Energy under contract number DE-AC02-76SF00515 to SLAC National Accelerator Laboratory.
The authors acknowledge The Red Española de Supercomputación for granting computing time to complete The300 project in the Marenostrum machine at the Barcelona Supercomputing Center.
This work has been made possible by \thethreehundred{} collaboration.\footnote{https://www.the300-project.org}
\section*{Data Availability}

Chandra X-ray data are available from the  Chandra Data Archive (CDA) at \url{https://cxc.harvard.edu/cda/}.
\citet{Mantz22} lists the specific observation IDs employed in that work, which are identical to those employed here.

\bibliographystyle{mnras}  
\bibliography{references}

\appendix

\section{SPA Metrics from Simulated Clusters} \label{sec:appendix}

\subsection{Calculations}
\label{sec:spametrics}

\begin{figure*}
\centering
\includegraphics[width=0.8\columnwidth]{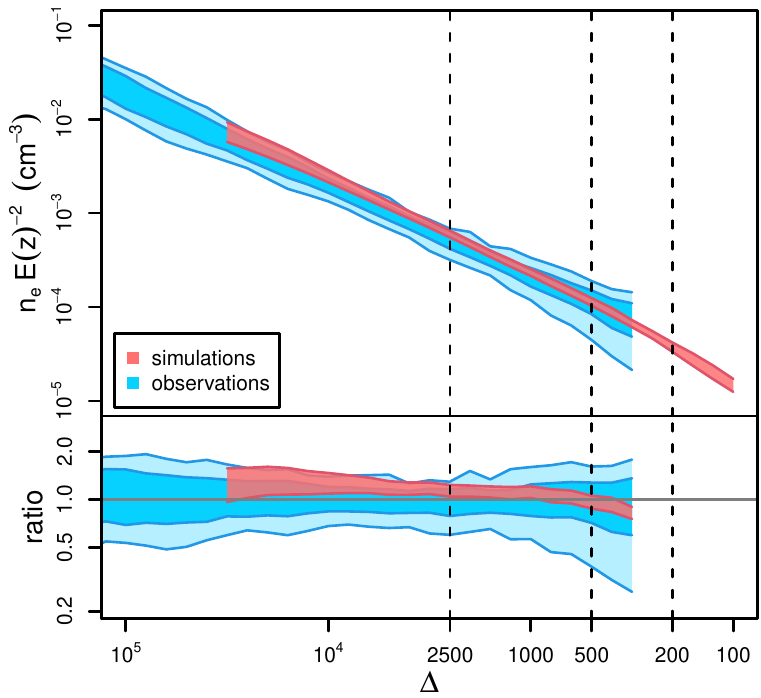}
\includegraphics[width=1.2\columnwidth]{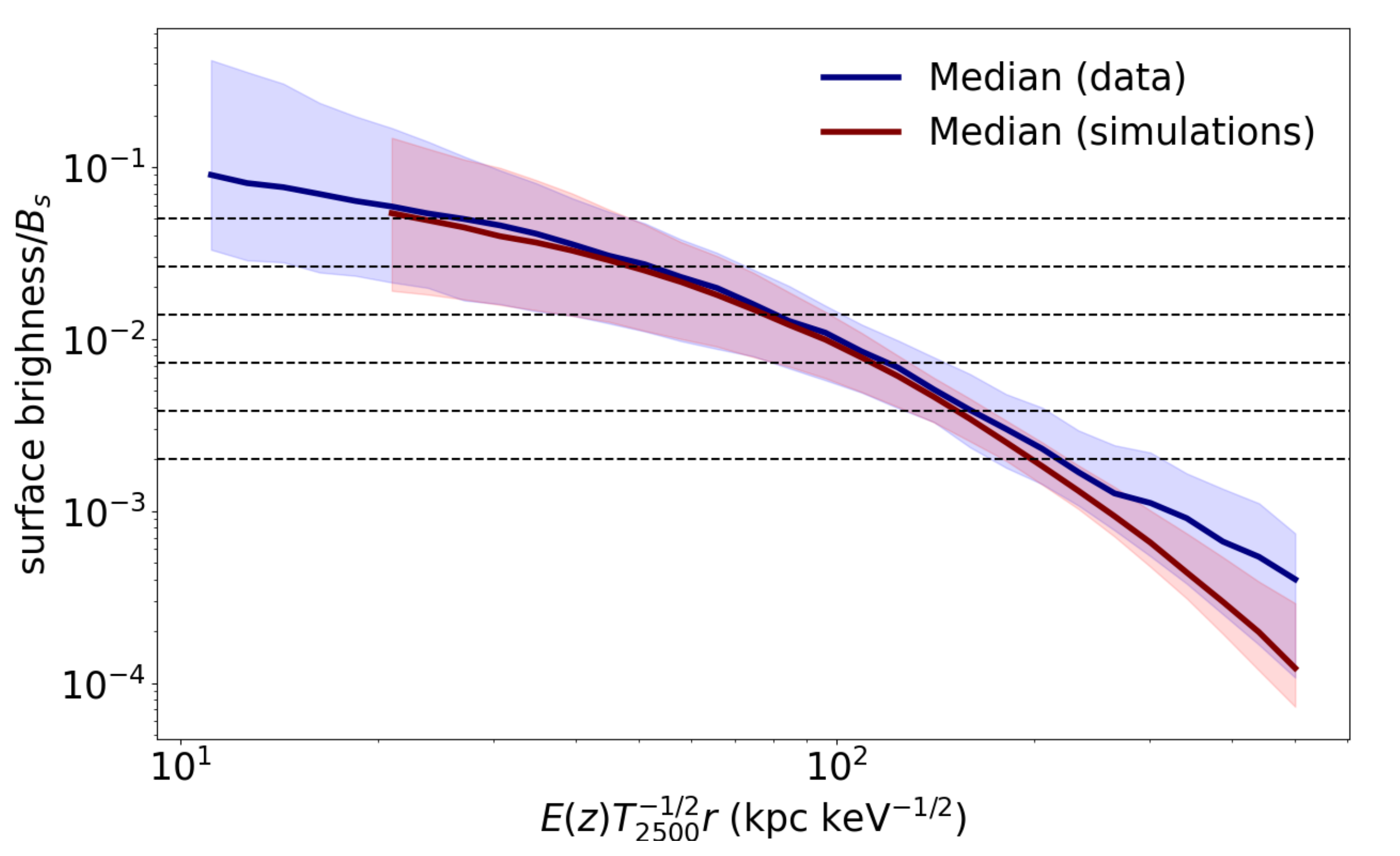}

\caption{\emph{Left}: Self-similarly scaled electron density profiles from {\it Chandra} observations of relaxed clusters (reproduced from \citealt{Mantz16}, with the Hubble constant adjusted for consistency with the simulations) and simulated clusters satisfying the same morphological selection. Plotting versus the enclosed overdensity, $\Delta$, rather than a simple scaled radius (e.g.\ $r/r_{500}$) reduces the scatter due to variations in cluster concentration. Blue shading shows the 68\% and 95\% confidence region from observations, including both statistical uncertainty and intrinsic scatter; red shading shows the 68\% region from intrinsic scatter in the simulations. Simulated results are limited to the range of overdensities where most clusters provide data, after excising the central 50\,kpc radius. The bottom-left panel shows the ratio to the median observed profile. \emph{Right:} Comparison of surface brightness between simulated clusters at z = 0.067 (\emph{red}) and observed clusters (\emph{blue}; M15) after accounting for self-similar scaling (cf Figure~2 of M15) and adjusting $B_S$ as described in the text. The dashed black levels delineate the five scaled isophotes used in the SPA analysis. Simulated results are limited to the range of overdensities where most clusters provide data, after excising the central 50\,kpc radius.}
\label{fig:sbplot}
\end{figure*}

The raw simulated X-ray maps encode the projected (line-of-sight integrated) X-ray emissivity in the 0.5--2.0\,keV band in the cluster rest frame (see \citealt{Ansarifard1911.07878} for details).
We convert these to maps of mean count rate in the observer-frame 0.6--2.0\,keV band (matching the M15 analysis) using the {\sc apec} model in {\sc xspec}, assuming a metallicity of 0.3 Solar (according to the Solar abundance table of \citealt{Asplund0909.0948}) and using the gas temperature at $r_{2500}$.
While the SPA algorithm accounts for the Poisson nature of X-ray data, it can be straightforwardly applied to the mean count rate maps, effectively assuming infinite signal-to-noise.
In order to derive observational metrics related as closely as possible to the target simulated cluster, we masked objects in the images that could be unambiguously identified as interlopers, i.e.\ those appearing in only one projection, since this requires them to be $>4$\,Mpc distant from the cluster center (Section~\ref{sec:sims}).
Such masking is only needed for $\sim 20$ of the $\sim900$ projections at each redshift.

The SPA metrics are computed by comparing these brightness maps to standard, self-similarly scaled surface brightness levels.
The scaling is given by\footnote{We have changed notation compared with M15 in order to avoid confusion with $f_s$, the the fraction of mass in subhalos (Section~\ref{sec:appendix_results}).}
\begin{equation} \label{eq:sbscal}
  B_S = \frac{K(z,T,N_\mathrm{H})}{5.7\times10^8} \frac{E(z)^3}{(1+z)^4} \left(\frac{kT}{\mathrm{keV}}\right) \frac{\mathrm{erg}}{\mathrm{Ms}\,\mathrm{cm}^2\,(0.984'')^2}.
\end{equation}
Here we depart from M15 by defining $K$ as the conversion factor between bolometric flux in energy units to photon flux (as opposed to energy flux) in the observed energy band.
This definition is more appropriate for cooler clusters ($kT\ltsim2$\,keV, given the adopted energy band), for which the energy-to-photon flux ratio varies with temperature; the renormalizing factor of $5.7\times10^8$ makes the definition approximately equivalent for the hotter clusters studied by M15.
The particular brightness levels used in the SPA algorithm are
\begin{equation} \label{eq:sblevels}
  S_j = 0.002 \times 10^{0.28j} B_S,
\end{equation}
where $j=0,1,\ldots,5$.

The motivation for approaching morphology measurement this way is that it allows comparable regions (in terms of enclosed overdensity) of different clusters to be identified based only on observed brightness, given intrinsic self-similarity of the gas density profiles at the relevant radii (outside the core), a roughly self-similar temperature--mass relation (used in deriving Equation~\ref{eq:sbscal}), and the relatively weak cosmological assumptions implicit in the normalized Hubble parameter, $E(z)$.
M15 showed that this procedure produces low-scatter scaled surface brightness profiles, and that the isophotes delineated by Equation~\ref{eq:sblevels} correspond to approximately comparable radii in units of $r_{2500}$, without requiring strong assumptions needed to estimate cluster masses (and thus characteristic radii such as $r_{2500}$) at the outset.
However, the regions identified in this way in observed clusters will only be comparable to those identified in simulated clusters if the distribution of gas in the simulated and observed clusters is similar at the radii of interest.
In particular, a difference in the gas mass fraction would straightforwardly lead to a difference in the surface brightness associated with a given overdensity, and disproportionately so due to the nonlinear dependence of X-ray emissivity on gas density.

We do see evidence for just such a difference between the simulations and the observed data.
This can be seen in the left panel of Figure~\ref{fig:sbplot}, which compares the measured density profiles for relaxed (SPA-selected) clusters from \citet{Mantz16} with the simulated clusters that meet the SPA criteria in at least 2 projections as a function of overdensity.\footnote{The selection of relaxed simulated clusters in this figure is made a posteriori in order to reduce the scatter at large radii due to merging sub-halos, thus visually clarifying the average difference between simulated and observed profiles, but the difference is present irrespective of the selection.}
There is a clear systematic offset at overdensities $>1000$, with the profiles coming into better agreement at $\Delta\sim500$; qualitatively similar behavior was seen by \citet{Li20}, comparing the same simulations (without the restriction to relaxed clusters) to measurements from \citet{McDonald17} (see also \cite{Campitiello2022, Sayers2022}).

As expected, this translates to an offset in the surface brightness profiles between the simulations and observations in appropriately scaled units, such that directly applying the equations above would mean assessing the morphology of a different region of clusters than the observations we compare to.
We therefore adjust $B_S$ by a factor of 2.2 when applied to the simulations, making the scaled surface brightness profiles agree well with the observed data, outside the smallest radii, which are most sensitive to the AGN feedback prescription (right panel of Figure~\ref{fig:sbplot}); as expected, this results in the radii identified by the isophotes (see below) resembling those identified in the real data.

With this preprocessing completed, the SPA calculations proceed as described by M15, with the exception that the simulated maps do not require estimation or propagation of measurement uncertainties. 
Briefly, peakiness ($p$) is defined as the average brightness in the circular region bounded by the radius where the cluster's surface brightness profile exceeds $S_5$ (Equation~\ref{eq:sblevels}), weighted by a factor of $1+z$ (Equation~5 of M15).
Symmetry ($s$) and alignment ($a$) depend on the centers of elliptical shapes fitted to the distribution of pixels falling in the ranges $S_0$--$S_1$, $S_1$--$S_2$, etc.: $s$ measures how well the centers of these ellipses agree with a global center, and $a$ measures how well they agree with one another (Equations~6--7 of M15).
Clusters with particularly faint centers may never reach the surface brightness thresholds corresponding to the brighter of these isophotes (this can occur in both the simulated and real data), and only those isophotes for which an ellipse can be fit contribute to $s$ and $a$.

\begin{figure*}
\centering
\includegraphics[width=0.32\textwidth]{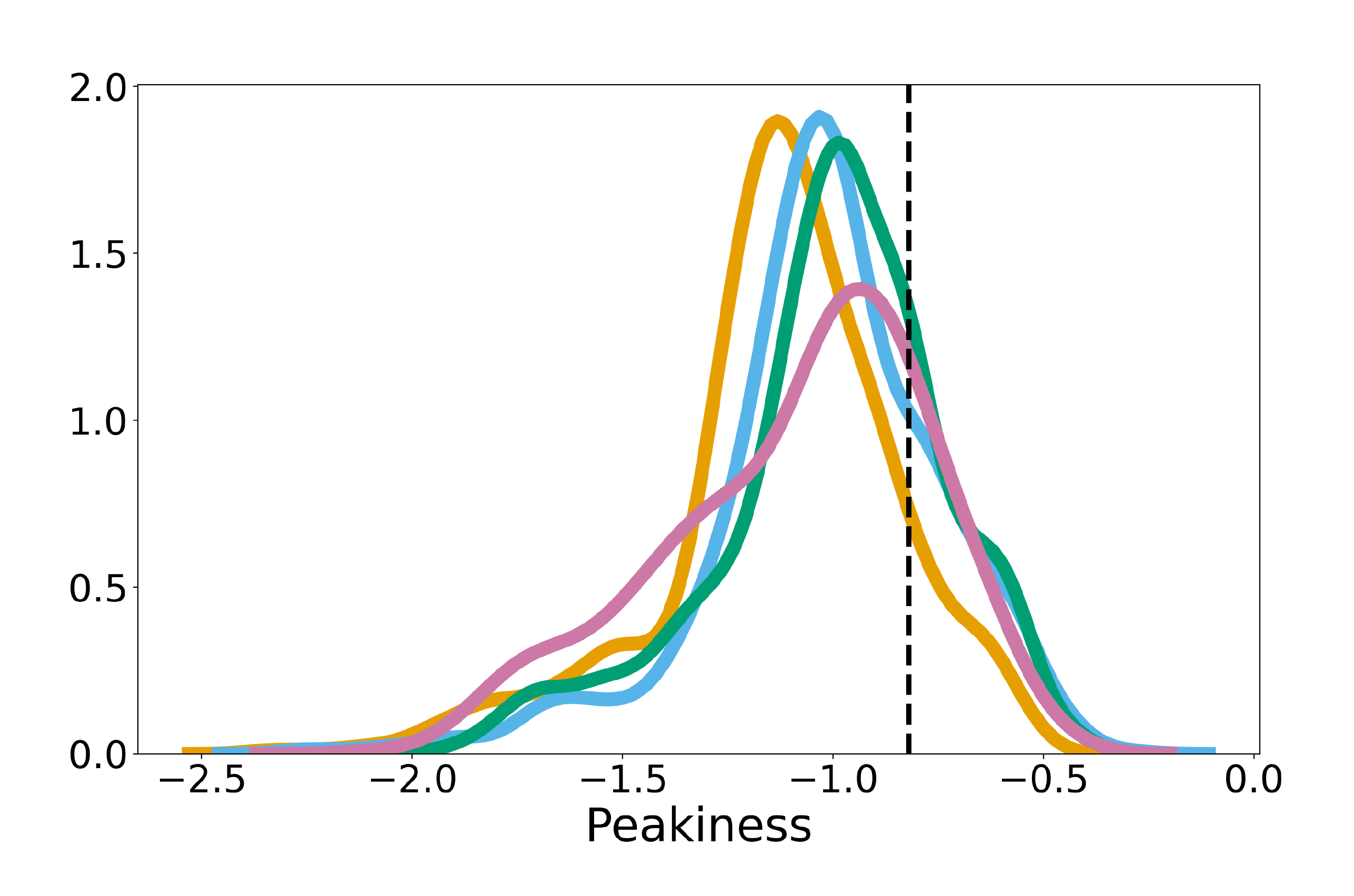}
\includegraphics[width=0.32\textwidth]{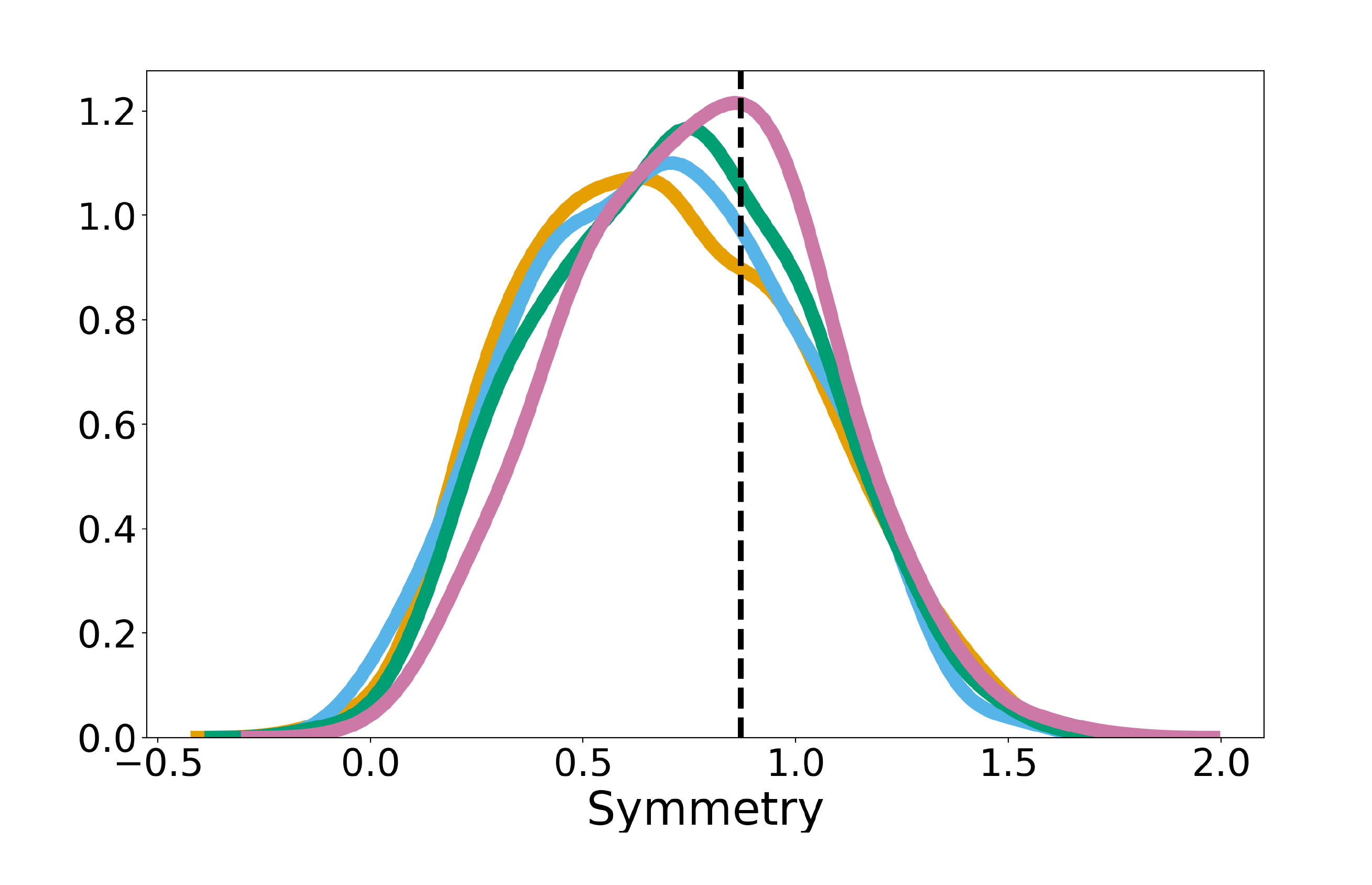}
\includegraphics[width=0.32\textwidth]{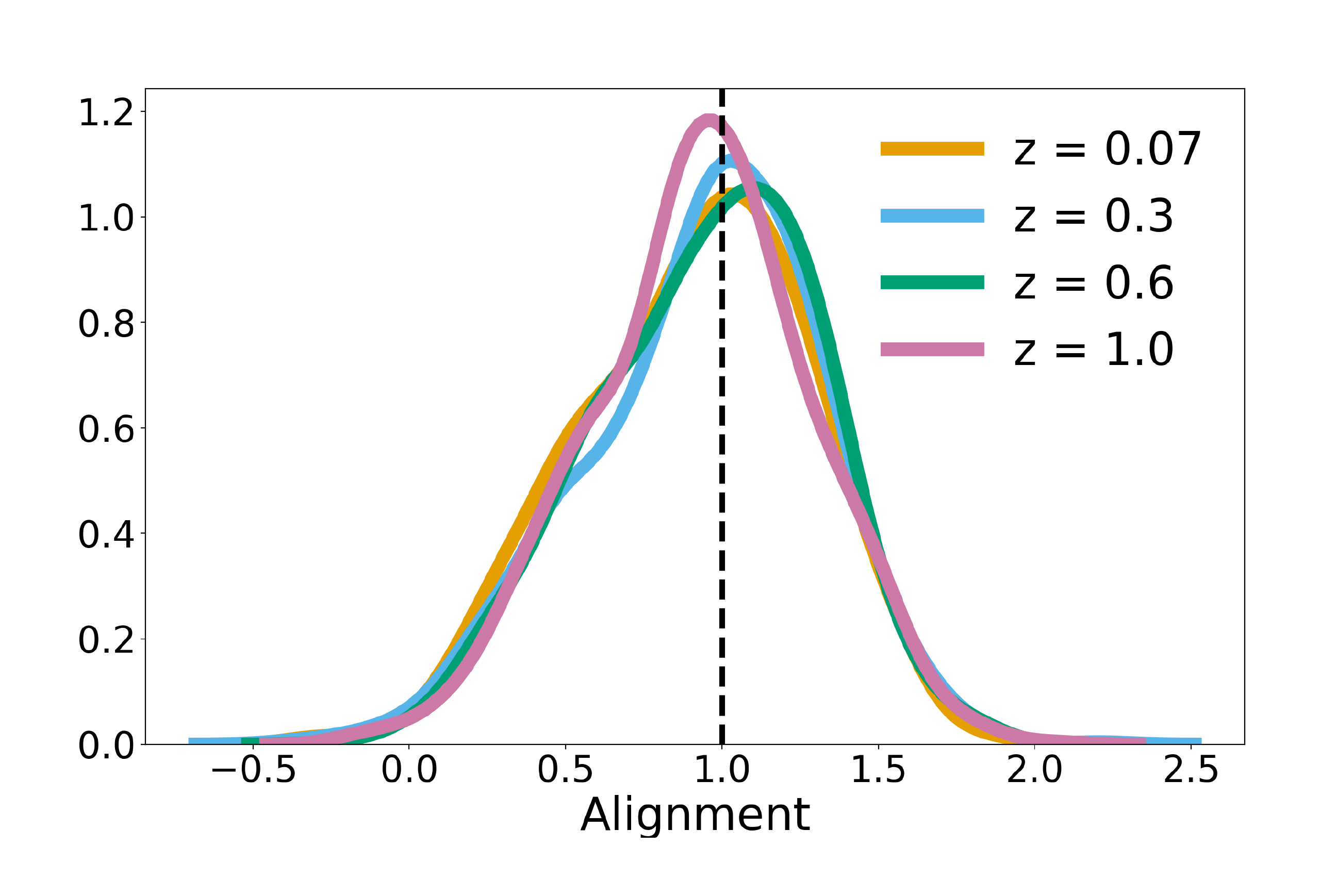}
\includegraphics[width=0.33\textwidth]{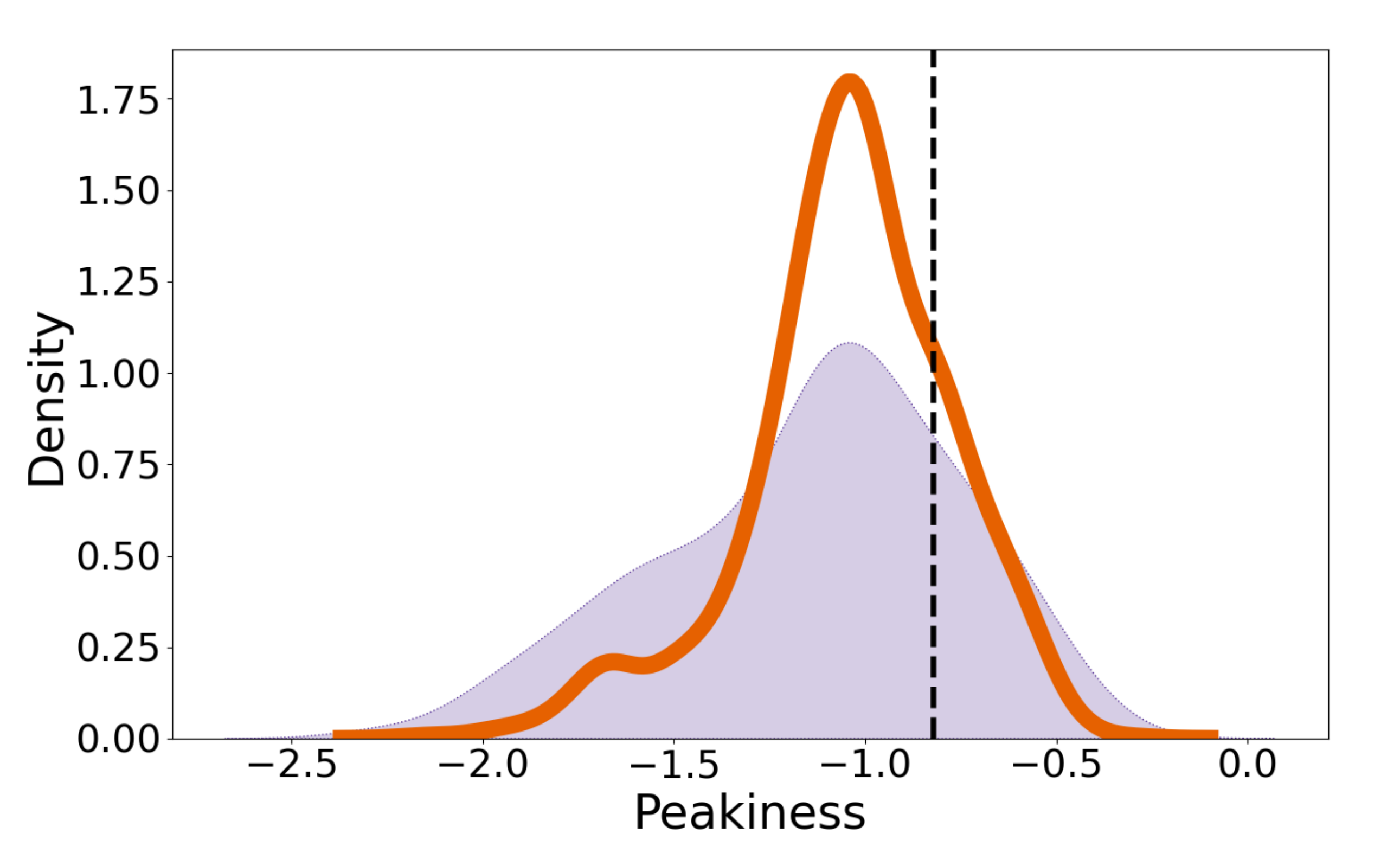}
\includegraphics[width=0.325\textwidth]{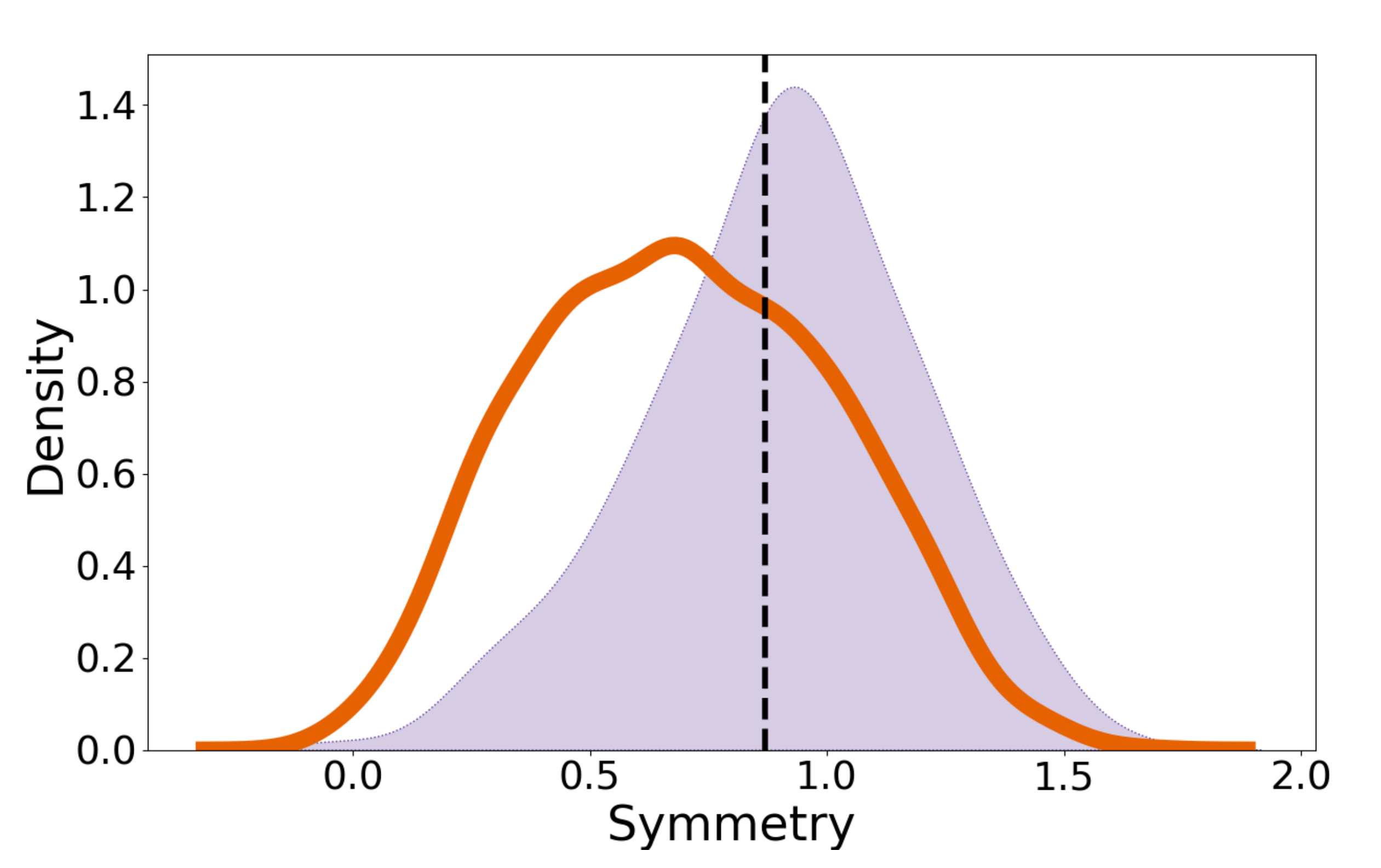}
\includegraphics[width=0.32\textwidth]{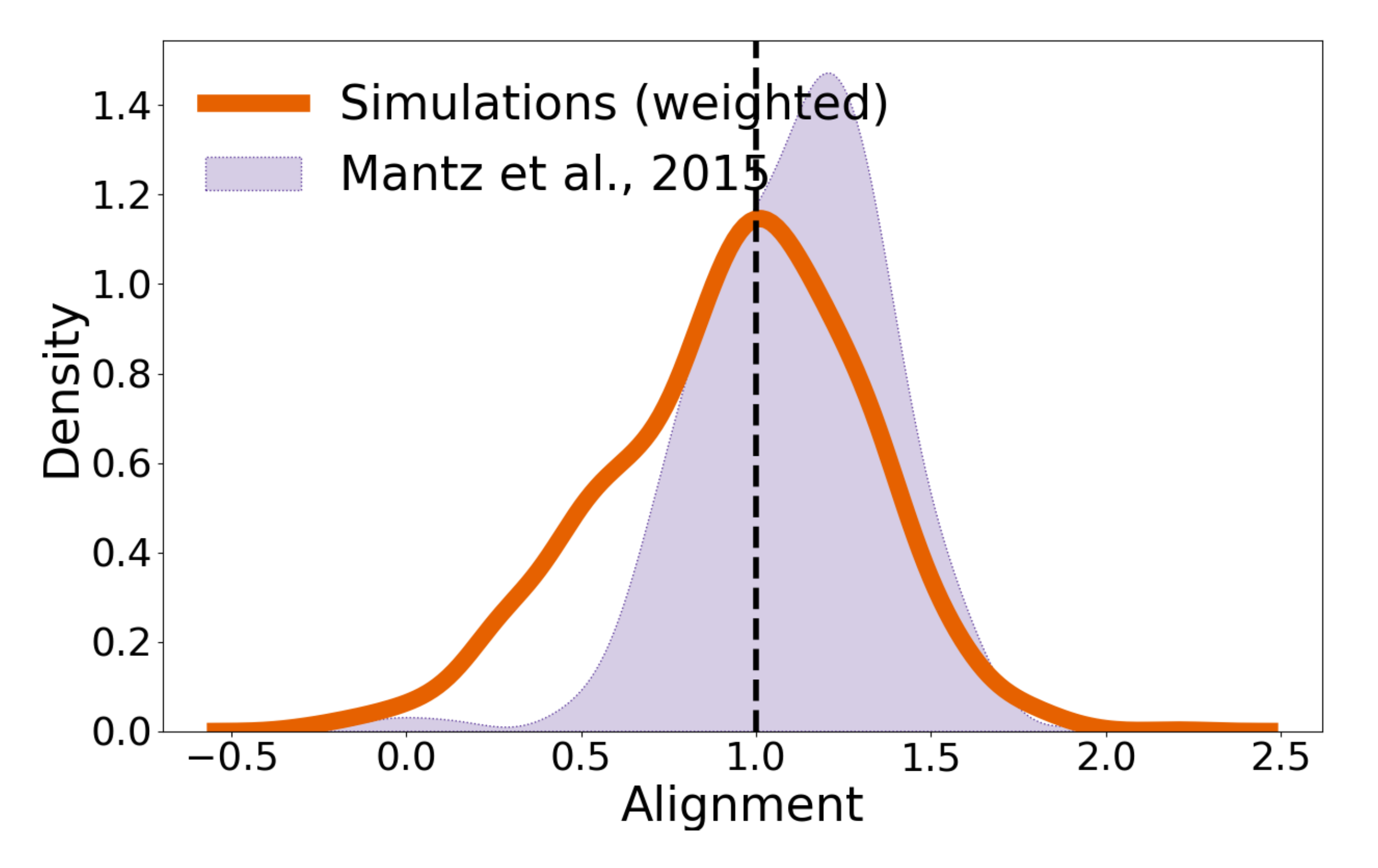}

\caption{Plot of $s$, $p$, and $a$ distributions for simulated clusters. \emph{top:} KDE plots for distributions split by redshift. \emph{bottom:} KDE plot for weighted combination of redshifts. Weights were chosen to approximate the same redshift distribution as the data used in the M15 analysis.}
\label{fig:SPA}
\end{figure*}

\begin{figure*}
\centering
\includegraphics[width=.51\textwidth]{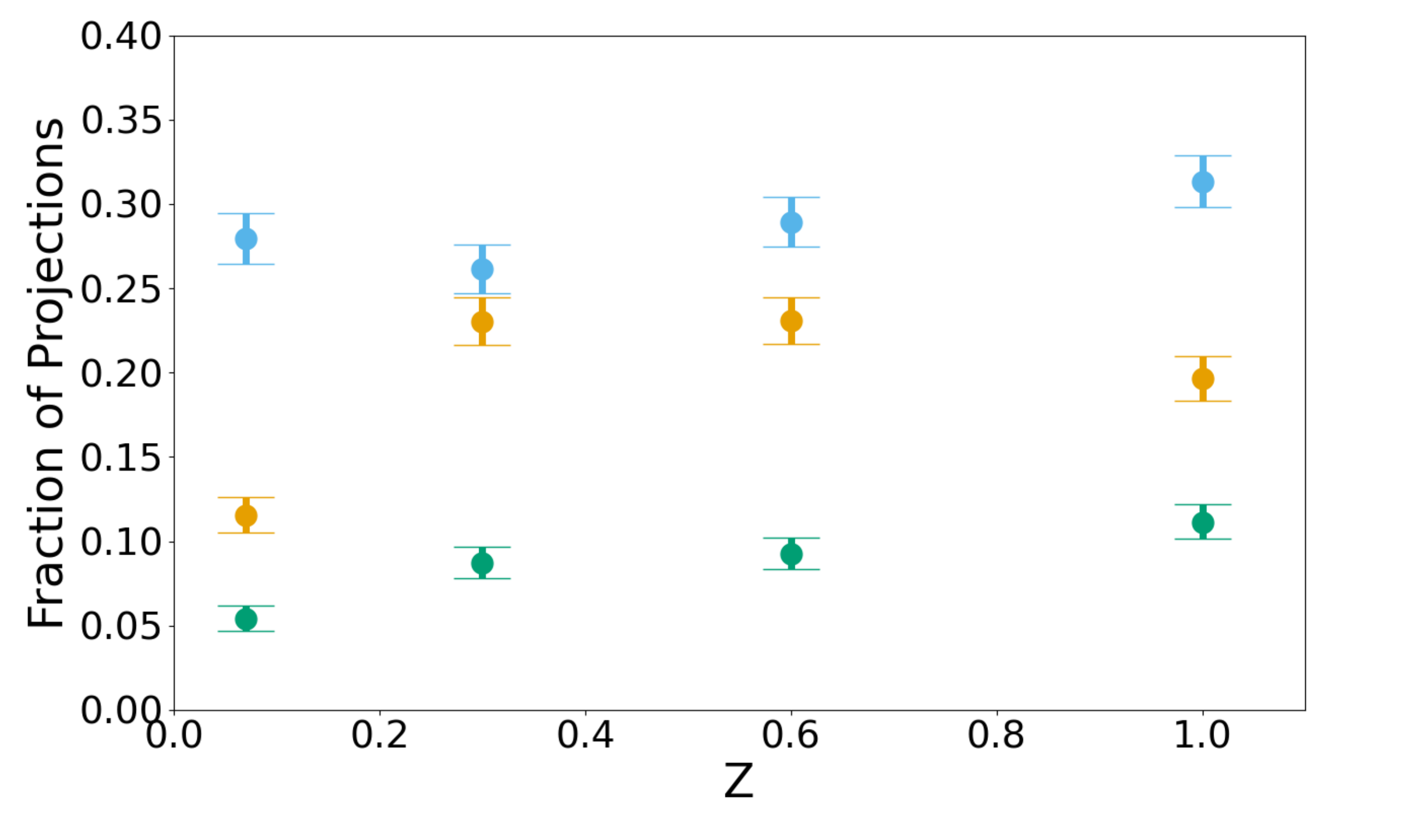}
\includegraphics[width=.48\textwidth]{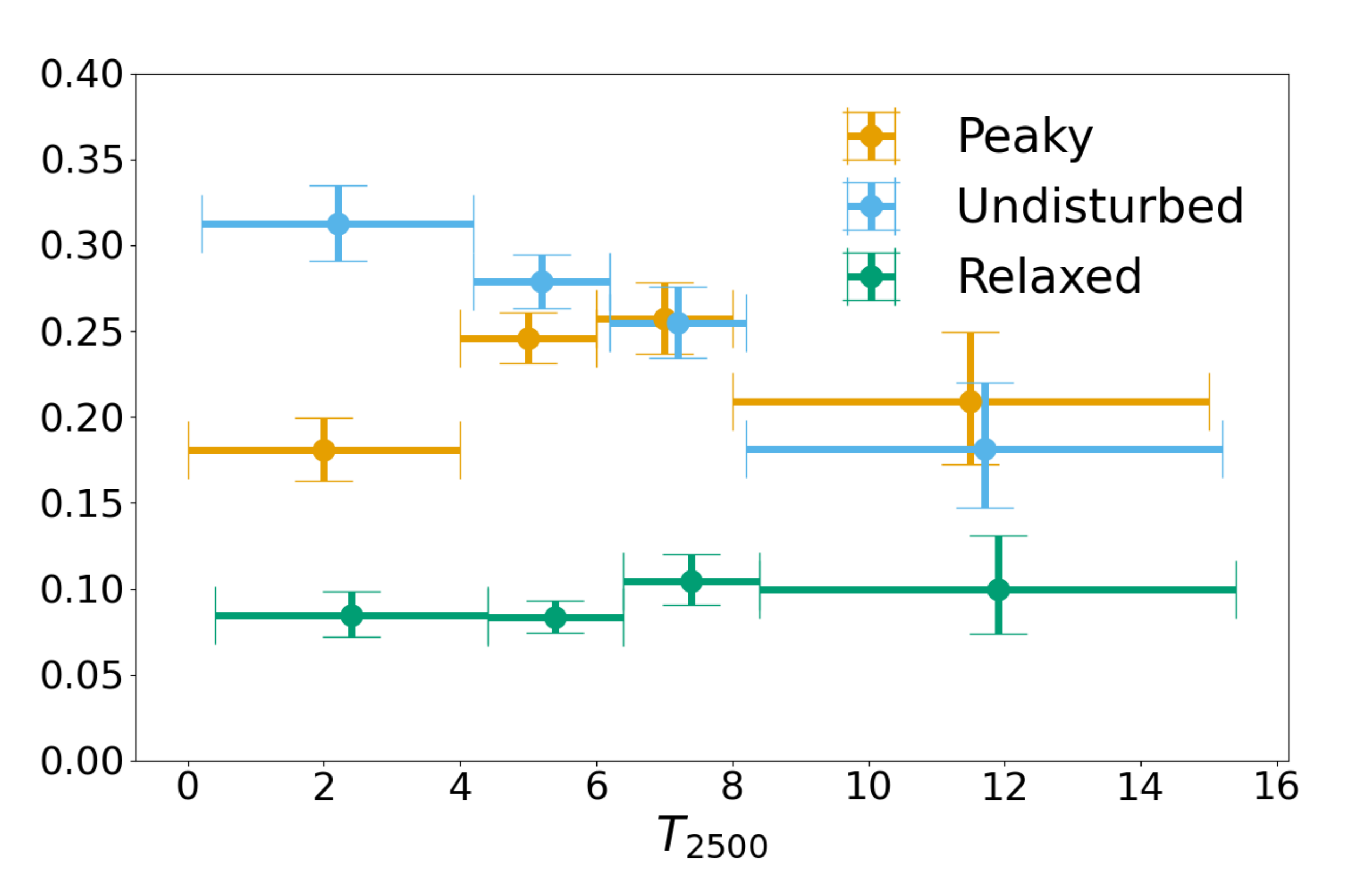}
\caption{The fraction of the projections for our simulated cluster sample identified as relaxed (\emph{green}), peaky (\emph{orange}), and undisturbed (\emph{blue}) as a function of redshift (\emph{left}) and mass-weighted temperature within $r_{2500}$ (\emph{right}). The y--errors represent the 68\% confidence interval. \emph{right:} the x--errors represent the bin sizes in temperature. Only $z = 0.3, 0.6$ are included in the temperature plot to control for the correlation between redshift and temperature. } 
\label{fig:relaxed_fraction}
\end{figure*}

\subsection{Comparison with Observed Clusters}
\label{sec:appendix_results} 

SPA values were calculated for each of the cluster projections, which we consider independently unless otherwise noted. The distributions of the SPA metrics for the simulated clusters can be seen in Figure \ref{fig:SPA}. The top row shows the distributions of parameters at each redshift separately. The bottom row compares a combination, approximately weighted by the redshift distribution of the M15 clusters, with the observed distributions from M15 ($z \in [0.08, 1.1]$).
The simulated distributions of $s$ and $a$ are somewhat broader than those measured from data, and offset to smaller values.
We see hints of evolution in the mode of the $s$ distributions, but otherwise the $s$ and $a$ distributions are quite consistent in shape across redshifts.

With $p$, we see a mild but consistent trend towards higher values in the mode and a larger spread with increasing redshift. The low $p$ tail at $z = 1.0$ is primarily driven by clusters with $kT < 5$ keV (Figure \ref{fig:relaxed_fraction}).  
Nevertheless, the upper tail of the peakiness distribution, exceeding the SPA selection threshold, is in good agreement with the data; it is also remarkably consistent for the simulated clusters across redshift, except for the lowest-redshift slice.

We apply the criterion for relaxation defined by M15, and also define ``peaky'' and ``undisturbed'' selections analogous to that work. The fraction of the sample identified as relaxed ($s > 0.87, p > -0.82$ and $a > 1.0$), peaky ($p > -0.82$), and undisturbed ($s > 0.87$ and $a > 1.0$) as a function of redshift can be seen in Figure \ref{fig:relaxed_fraction} (cf.\ Figure~\ref{fig:SPA}). 
We see a slight trend towards increasing relaxed fraction with increasing redshift. The fraction of relaxed clusters for $z>0.3$ is roughly consistent with the $11\%$ of observed clusters classified as relaxed by M15. 

Comparing with Figure 18 of M15, we see that the relaxed, peaky, and undisturbed fractions are in good agreement with their SZ-selected cluster samples (respectively: 8.5\%, 14\%, and $\sim$ 20\%), tending to be lower on average than the measured fractions for their X-ray selected samples. In total, 11\% of the observed and 8\% of the simulated sample is identified as relaxed. This corresponds to 44 observed and 96 simulated clusters (15 at $z=0.07$, 23 at $z=0.3$, 28 at $z=0.6$, and 30 at $z=1.0$) included in the relaxed sample.

\begin{table*}
\centering
\caption{Criteria for cluster relaxation taken from previous works. (1) gives the cuts used to select the sample of simulated clusters (2) is the total fraction of the sample selected as relaxed across all redshifts (3) gives the mean and standard deviation of $\Delta z_\mathrm{form}$ for the relaxed sample, and (4-5) show the completeness and purity of the sample relative to the SPA selection.}
\label{tab:relaxed_frac}
\begin{tabular}{lcccr} 
\hline
Selection Criteria & Selected Fraction  &  $\Delta z_\mathrm{form}$ & Completeness   & Purity \\ 
\hline
$0.85 < \eta_{2500} < 1.15, \Delta r_{2500} < 0.04$, $f_{s,2500} < 0.1$ \citep{Cui_2018} & 4.5\%  & $0.63 \pm 0.29 $ & 9.2\% & 17.7\% \\
$\Delta r_{500} < 0.1$, $f_{s,500} < 0.1$ \citep{De_Luca_2021} & 43.3\% & $0.62 \pm 0.30 $ & 78.5\% & 15.7\% \\
$\Delta r_{500} < 0.07$, $f_{s,500} < 0.1$ \citep{Cao_2021} & 40.2\% & $0.63 \pm 0.30 $ & 77.2\% & 16.7\% \\
$s > 0.87, a > 1.0, p > -0.82$ (M15) & 8.7\% & $0.75 \pm 0.28$ & -- & -- \\
\hline
\end{tabular}

\end{table*}

We also show a comparison between the SPA criteria for relaxation and selection criteria for simulated clusters based on the virial ratio ($\eta$), the center-of-mass offset ($\Delta r$) and the fraction of mass in subhalos ($f_s$) in Table \ref{tab:relaxed_frac}. These metrics are commonly used in studies of simulated clusters as simply measurable proxies for relaxation \citep{Cui_2018, De_Luca_2021, Cao_2021}.  As can be seen, the criteria used in previous works bracket the selectivity of the SPA cuts, with the selection criteria from \cite{Cui_2018} identifying only $4.5\%$ of clusters as relaxed with limited overlap, with the SPA--selected sample. On the other end, the \cite{De_Luca_2021} cuts select $\sim 40\%$ of clusters as relaxed and include almost 80\% of the SPA selected clusters. While all selected samples show large scatter in $\Delta z_\mathrm{form}$, the cuts based on the SPA relaxed sample show a higher value of $\Delta z_\mathrm{form}$ than the relaxed samples defined using previous cuts.

\bsp	
\label{lastpage}
\end{document}